\documentclass[12pt,preprintnumbers,nofootinbib,letterpaper]{article}
\pdfoutput=1
\usepackage{amsmath,amssymb,amsfonts,cite,xcolor,epsf,epsfig,epstopdf,graphics,graphicx,mathtools,subcaption,physics,verbatim}
\def\thefootnote{*\arabic{footnote}}
\definecolor{ultramarine}{rgb}{0.07, 0.04, 0.56}
\definecolor{cadmiumgreen}{rgb}{0.0, 0.42, 0.24}
\definecolor{indigo(dye)}{rgb}{0.0, 0.25, 0.42}
\usepackage[linktocpage=true,breaklinks]{hyperref}
\hypersetup{
	colorlinks=true,
	citecolor=ultramarine,
	linkcolor=cadmiumgreen,
	urlcolor=indigo(dye),
}

\usepackage[utf8]{inputenc}

\numberwithin{equation}{section}

\usepackage{array}
\newcolumntype{P}[1]{>{\centering\arraybackslash}p{#1}}

\usepackage[shortlabels]{enumitem}

\usepackage{dcolumn}
\newcolumntype{M}[1]{>{\centering\arraybackslash}m{#1}}
\newcolumntype{N}{@{}m{0pt}@{}}

\usepackage{amsmath}
\usepackage{ulem}

\newcommand{\be}{\begin{equation}}  
\newcommand{\ee}{\end{equation}}

\usepackage{tikz}
\begin{document}


\begin{center}

\def\thefootnote{\fnsymbol{footnote}}

\begin{flushright} {\footnotesize YITP-25-130, IPMU25-0043 }  \end{flushright}

\vspace*{1.5cm}
{\Large {\bf Linear Higher-Order Maxwell-Einstein-Scalar Theories}}
\\[1cm]

{Mohammad Ali Gorji$^{1}$,
Shinji Mukohyama$^{2,3}$,
Pavel Petrov$^{1}$,
Masahide Yamaguchi$^{1,4,5}$}
\\[.7cm]

{\small \textit{$^1$Cosmology, Gravity, and Astroparticle Physics Group, Center for Theoretical Physics of the Universe, Institute for Basic Science (IBS), Daejeon, 34126, Korea
}}\\*[7pt]

{\small \textit{$^{2}$Center for Gravitational Physics and Quantum Information,
\\Yukawa Institute for Theoretical Physics, Kyoto University, 606-8502, Kyoto, Japan}}\\*[7pt]
	
{\small \textit{$^{3}$Kavli Institute for the Physics and Mathematics of the Universe (WPI), The University of Tokyo Institutes for Advanced Study, The University of Tokyo, Kashiwa, Chiba 277-8583, Japan}}\\*[7pt]

{\small \textit{$^{4}$Department of Physics, Institute of Science Tokyo, 
\\2-12-1 Ookayama, Meguro-ku, Tokyo 152-8551, Japan}}\\*[7pt]

{\small \textit{$^{5}$Department of Physics and IPAP, Yonsei University, 50 Yonsei-ro, Seodaemun-gu, Seoul 03722, Korea}}\\

\end{center}

\vspace{1cm}

\hrule \vspace{0.5cm}

\begin{abstract}
In the context of the Higher-Order Maxwell–Einstein–Scalar (HOMES) theories, which are invariant under spacetime diffeomorphisms and $U(1)$ gauge symmetry, we study two broad subclasses: the first is up to linear in $R_{\mu\nu\alpha\beta}$, $\nabla_\mu\nabla_\nu\phi$, $\nabla_\rho{F}_{\mu\nu}$ and up to quadratic in the vector field strength tensor $F_{\mu\nu}$; the second is up to linear in $\nabla_\mu\nabla_\nu\phi$, contains no second derivatives of vector field and metric, but allows for arbitrary functions/powers of $F_{\mu\nu}$. Under these assumptions, we systematically derive the most general form of the action that leads to second-order (or lower) equations of motion. We prove that, among 41 possible terms in the first subclass, only four independent higher-derivative terms are allowed: the kinetic gravity braiding term $G_3(\phi,X)\Box\phi$ in the scalar sector with $X = -\nabla_\mu\phi \nabla^\mu\phi / 2$; the Horndeski non-minimal coupling term $w_0(\phi)R_{\beta \delta \alpha \gamma}\tilde{F}^{\alpha \beta } \tilde{F}^{\gamma \delta }$ in the vector field sector, where $\tilde{F}^{\mu\nu}$ is the Hodge dual of $F_{\mu\nu}$; and two interaction terms between the scalar and vector field sectors: $[w_1(\phi,X) g_{\rho\sigma} + w_2(\phi,X) \nabla_{\rho}\phi \nabla_{\sigma}\phi] \nabla_\beta\nabla_\alpha\phi \, \tilde{F}^{\alpha \rho } \tilde{F}^{\beta\sigma}$. For the second subclass, which admits 11 possible terms, three of these four, excluding the Horndeski non-minimal coupling term proportional to $w_0(\phi)$, are allowed. These independent terms serve as the building blocks of each subclass of HOMES. Remarkably, there is no higher-derivative parity-violating term in either subclass. Finally, we propose a \textit{new} generalization of higher-derivative interaction terms for the case of a charged complex scalar field.
\end{abstract}
\vspace{0.5cm} 

\hrule
\def\thefootnote{\arabic{footnote}}
\setcounter{footnote}{0}

\thispagestyle{empty}


\newpage

\section{Introduction and summary}
\label{sec: Introduction}

Scalar-tensor and vector-tensor theories with higher derivatives in the action have been widely studied in recent years. Taking higher derivatives into account requires careful consideration, as extra ghost degree(s) of freedom generally appear. There are different ways to address this issue. One approach is to impose second-order equations of motion to ensure the correct number of propagating degrees of freedom for the system under consideration. Another method involves imposing the so-called degeneracy condition on the action, which allows for higher-order (beyond second order) equations of motion while still maintaining the correct number of propagating degrees of freedom\footnote{The first approach can be regarded as a part of the second approach because theories obtained in the first approach must be degenerate.}. Yet another approach is based on effective field theory methods, in which the extra (would-be ghost) degrees of freedom are heavy and can be safely integrated out, leaving a system with the correct propagating degrees of freedom. These approaches have been employed to study various scalar-tensor and vector-tensor theories. 

In the case of scalar-tensor theories, the most general theory with second-order equations of motion is Horndeski gravity~\cite{Horndeski:1974wa} (see Refs.\cite{Rubakov:2014jja, Babichev:2016rlq, Kobayashi:2019hrl, Mironov:2024pjt} for reviews), which was shown in Ref.\cite{Kobayashi:2011nu} to be equivalent to the generalized Galileons~\cite{Deffayet:2011gz}. On the other hand, the so-called degenerate Higher-Order Scalar-Tensor (DHOST) theories extend the framework to include higher-order (beyond second-order) equations of motion while still maintaining the correct number of propagating degrees of freedom~\cite{Langlois:2015cwa,BenAchour:2016cay,BenAchour:2016fzp,BenAchour:2024hbg}. The effective field theory approach to scalar-tensor theories has been developed in Refs.\cite{Arkani-Hamed:2003pdi,Cheung:2007st,Gubitosi:2012hu}, providing a systematic way to incorporate higher-derivative terms in a healthy manner\cite{Mukohyama:2005rw,Mukohyama:2006be,Motohashi:2019ymr,Gorji:2020bfl,Gorji:2021isn,DeFelice:2022xvq}. 

For vector-tensor theories, ghost-free theories with five propagating degrees of freedom have been formulated which is known as generalized Proca theories~\cite{Tasinato:2014eka,Heisenberg:2014rta,Allys:2015sht,Heisenberg:2016eld}. The effective field theory of vector-tensor theories has been developed in Refs.~\cite{Mukohyama:2006mm,Aoki:2021wew,Aoki:2023bmz}. In the context of generalized Proca theories, the vector field does not respect the $U(1)$ symmetry, and the resulting action does not contain higher-order derivatives of the vector field itself. Additionally, a model was proposed that includes second-order derivatives of the vector field in the action while still yielding second-order field equations~\cite{Petrov:2018xtx, Petrov:2020vlq}.  However, this model also does not preserve gauge invariance for the vector field. Therefore, neither  class of vector field models \textit{can} be considered Horndeski-type generalizations of electromagnetism. Interestingly, the absence of $U(1)$ symmetry in these models is not coincidental. As shown in Ref.~\cite{Deffayet:2013tca} for a single $U(1)$ field, Horndeski-type generalizations in flat spacetime in four dimensions are forbidden. There are three main known ways to overcome this kind of no-go theorem. The first approach is to relax the requirement of second-order equations of motion and instead require only the absence of Ostrogradsky ghosts~\cite{Colleaux:2025vtm}. Theories of this kind are natural vector field generalizations of DHOST theories \cite{Langlois:2015cwa, Langlois:2015skt}. The second approach is to consider terms in the action that identically vanish in flat spacetime~\cite{Deffayet:2013tca, Nandi:2017ajk}. Finally, the third approach was proposed in Ref.\cite{Deffayet:2010zh}. The essence of this idea is to introduce additional scalar fields. We refer to such models as Higher-Order Maxwell-Einstein–Scalar (HOMES) theories. However, only a few specific examples of these theories were constructed in Ref.\cite{Deffayet:2010zh, Heisenberg:2018acv}, and none of them are linear in the second derivatives of the vector field. We emphasize that HOMES theories are fundamentally different from generalized Proca models~\cite{Heisenberg:2014rta} even if the $U(1)$ gauge symmetry is restored in the latter by the introduction of the Stückelberg field, since the real scalar field in HOMES theories is invariant under the $U(1)$ gauge transformation while the Stückelberg field is not. Furthermore, in principle, HOMES theories contain second and (higher) order derivatives from the vector field unlike the Generalized Proca theories. Recently, by using Kaluza–Klein compactification of the 5D Horndeski theory, more examples of HOMES have been constructed~\cite{Mironov:2024idn , Mironov:2024umy }. However, the most general Lagrangian for HOMES models is still absent. Therefore, in this paper, we provide a complete analysis of two broad subclasses of HOMES theories and derive the most general form of the HOMES Lagrangian for each.

The {\it first subclass} is defined under the following assumptions:
\begin{enumerate}[label=\textbf{I-\arabic*.}]
	
	 \item The action respects spacetime diffeomorphisms and $U(1)$ gauge symmetry,
	 
    \item Up to linear order in $\{\phi_{\mu\nu},\nabla_\rho{F}_{\mu\nu}, R_{\mu\nu\alpha\beta}\}$ in the action,\footnote{It is worth mentioning that, as shown in Ref.~\cite{Mironov:2024umy}, all known examples of HOMES theories that include more than one second derivative of the vector field admit equal propagation speeds for photons and gravitons on FLRW backgrounds only when the Lagrangian functions are fine-tuned and a specific background solution is chosen. However, the GW170817 event (see Ref.~\cite{LIGOScientific:2017vwq}) imposes very stringent constraints on the speed of gravitational waves at late times. In this respect, the subclass of HOMES theories that are linear in second derivatives is phenomenologically more viable.}
    
    \item The equations of motion of all fields $\{\phi, A_\mu, g_{\mu\nu}\}$ are second-order,
    
    \item Up to quadratic order in vector field $A_\mu$,

\end{enumerate}
where $F_{\mu\nu} \equiv \nabla_{\mu} A_{\nu} -  \nabla_{\nu} A_{\mu}$ is the field strength tensor of $A_\mu$, $R_{\mu\nu\alpha\beta}$ is the Riemann tensor, and we have used the notation $\phi_{\mu} \equiv \nabla_\mu\phi$ such that $\phi_{\mu\nu}=\nabla_\nu\nabla_\mu\phi$. We show that, in general, this subclass includes 41 arbitrary functions of two variables as a starting point before imposing $\textbf{I-3}$. 

The {\it second subclass} is defined as:
\begin{enumerate}[label=\textbf{II-\arabic*.}]
	    
	\item The action respects spacetime diffeomorphisms and $U(1)$ gauge symmetry,
	
    \item Up to linear order in the second derivatives of the scalar field $\phi_{\mu\nu}$ in the action, but without second derivatives of the vector field and/or the metric,
    
    \item The equations of motion of all fields $\{\phi, A_\mu, g_{\mu\nu}\}$ are second-order.

\end{enumerate}
We show that this subclass includes 11 arbitrary functions of two variables as a starting point before imposing $\textbf{II-3}$.

Imposing the above conditions and performing the full analysis, we arrive at the following Lagrangian density for the first subclass and that without $w_0$ for the second subclass:
\begin{align}\label{L-HOMES}
\begin{split}
L_{\rm H} &= \text{lower-order part} - G_3(\phi,X)\,\square\phi 
+ \big[
w_0(\phi)\,R_{ \beta \delta \alpha \gamma}
+ (w_1(\phi,X) g_{\beta\delta} 
\\
&+ w_2(\phi,X)\phi_{\beta}\phi_{\delta}) \phi_{\gamma\alpha} 
\big] \tilde{F}^{\alpha \beta } \tilde{F}^{\gamma \delta } \;,
\end{split}
\end{align}
where $G_3$ and $w_i$ are arbitrary functions, $X=-\phi_{\mu}\phi^{\mu}/2$, and $\tilde{F}^{\mu\nu}=\frac{1}{2}\epsilon^{\mu\nu\rho\sigma}F_{\rho\sigma}$ is the Hodge dual of 
$F_{\mu\nu}$ with $\epsilon^{\mu\nu\rho\sigma}$ being the Levi-Civita tensor. Note that under our assumptions we have \textit{no} higher-order terms in parity-odd sector. The $G_3$ term is clearly the so-called Kinetic Gravity Braiding (KGB) term in the scalar sector \cite{Deffayet:2010qz} while $w_0$ term corresponds to the well-known Horndeski non-minimal interaction in vector field sector \cite{Horndeski:1976gi}. As the Lagrangian above is linear in second derivatives of the fields $\{\phi, A_\mu, g_{\mu\nu}\}$, one may consider it as a natural generalization of the KGB  which includes a $U(1)$ vector field. It is interesting to note that for the mixing between vector field and higher-order derivatives of scalar field, among many possible terms, there are only two terms ($w_{1,2}$) which indeed coincide with $U(1)$ symmetric scalar-vector-tensor theories~\cite{Heisenberg:2018acv}\footnote{We have noticed that in Ref.~\cite{Heisenberg:2018acv}, the term involving $\tilde{f}_4$ leads to higher-order derivatives in the metric field
equations and hence should vanish. The construction of gauge-invariant 
scalar-vector-tensor theories in Ref.~\cite{Heisenberg:2018acv} is different from HOMES theories as emphasized in the third paragraph in this section. Indeed, HOMES theories reduce to gauge-invariant scalar-vector-tensor theories in the absence of second (or higher) derivatives of the vector field. We thank Lavinia Heisenberg for confirming this point.}. Note that in Ref. \cite{Heisenberg:2018acv}, terms which include $\nabla_\rho{F}_{\mu\nu}$ have not been included. We have shown that, even if we relax this condition, as far as the action is quadratic in $F_{\mu\nu}$, such possible terms are all redundant to $w_{1,2}$ terms. Therefore, we prove the following generic theorems:  
\begin{itemize}
    \item The \textit{most} general HOMES Lagrangian which satisfies conditions $\textbf{I-1}$ to $\textbf{I-4}$ is given by Eq.~\eqref{L-HOMES},
   
    \item The \textit{most} general HOMES Lagrangian which satisfies conditions $\textbf{II-1}$ to $\textbf{II-3}$ is given by Eq.~\eqref{L-HOMES}, with $w_0$ set to zero.
\end{itemize}

Finally, to go beyond the current setup while keeping the higher-order terms at the same order, we generalize HOMES Lagrangian to the case of $U(1)$ charged scalar field~\footnote{The higher-derivative couplings between the complex scalar field and the vector field were also discussed in Ref.~\cite{Hull:2014bga}.}. A \textit{new} example of this generalization is given by
\begin{align}\label{L-complex}
\begin{split}
L_C &=  s_1 (\rho,\chi) \big[ \phi (D^\mu D_\mu \phi)^* + \mbox{c.c.} \big] 
+ s_2 (\rho, \chi) \big( \phi^* D^{\mu}D^{\nu}\phi + \mbox{c.c.} \big) \tilde{F}_{\mu\alpha} \tilde{F}_{\nu}{}^{\alpha}  \\ 
&+ \big[ s_3(\rho, \chi)D^{\alpha}\phi (D^{\beta}\phi)^* + s_4(\rho,\chi)\big({\phi^{*}}^2 D^{\alpha}\phi D^{\beta}\phi + \mbox{c.c.} \big) \big] \big( \phi^* D^{\mu}D^{\nu}\phi + \mbox{c.c.} \big) \tilde{F}_{\mu\alpha}\tilde{F}_{\nu\beta}\,,
\end{split}
\end{align} 
where $\rho=\sqrt{\phi\phi^*}$, $\mbox{c.c.}$ denotes the complex conjugate, and we have defined 
\begin{align*}
D_{\mu} \equiv \nabla_{\mu} - i e A_{\mu} \,,
\qquad
\chi &\equiv - |\phi|^2 D^{\mu} \phi (D_{\mu} \phi)^*
 -\frac{1}{2} \big( {\phi^*}^2 D^{\mu} \phi D_{\mu} \phi +\mbox{c.c.} \big) \;.
\end{align*}
Note that $\chi$ includes specific combinations of the derivatives of $\phi$ which, otherwise, would lead to higher-derivative (beyond second-order) equation of motion. 

The rest of the paper is organized as follows. In Sec.~\ref{sec: The HOMES Lagrangian}, we explicitly show that how apparently different higher-derivative terms are subset of our four higher-order derivative terms in Lagrangian \eqref{L-HOMES} which shows that these terms serve as a basis for the linear HOMES models that are discussed in the literature. In Sec.~\ref{sec: Lagrangian construction for the first case}, we consider the first subclass and list all possible independent combinations that can appear in the HOMES Lagrangian. By imposing the above-mentioned conditions, we systematically prove that there are \textit{only} four independent higher-order derivative terms. In Sec.~\ref{sec: Lagrangian construction for the second case}, we follow the same procedure for the second subclass. Sec.~\ref{sec: Conclusion} is devoted to the summary and discussion.

\section{The HOMES Lagrangian}
\label{sec: The HOMES Lagrangian}

The most general parity-even Lagrangian of the first subclass, defined by conditions $\textbf{I-1}$ to $\textbf{I-4}$ in Sec.~\ref{sec: Introduction}, is
\begin{align}
\label{L-pe-EMS}
\begin{split}
L^{pe}_{\rm H} &= -\frac{1}{4}g_0 (\phi, X) F_{\mu\nu} F^{\mu\nu} + g_2(\phi,X)  F^{\rho\mu}F_{\rho}{}^{\nu}\phi_{\mu}\phi_{\nu} +  G_2 (\phi, X) - G_3(\phi,X)\,\square\phi 
\\
&+ \big[
w_0(\phi)\,R_{ \beta \delta \alpha \gamma}
+ (w_1(\phi,X) g_{\beta\delta} 
+ w_2(\phi,X) \phi_{\beta} \phi_{\delta}) \phi_{\alpha\gamma} 
\big] \tilde{F}^{\alpha \beta } \tilde{F}^{\gamma \delta } \;,
\end{split}
\end{align}
where $G_i$, $g_i$, and $w_i$ are arbitrary functions. For the second subclass, defined by conditions $\textbf{II-1}$ to $\textbf{II-3}$ in Sec.~\ref{sec: Introduction}, one obtains a subset of the above Lagrangian with $w_0=0$. A detailed proof for each subclass is given in Sec.~\ref{sec: Lagrangian construction for the first case} and Sec.~\ref{sec: Lagrangian construction for the second case}, respectively. One can directly check that the Lagrangian density \eqref{L-pe-EMS} generates second-order equations for all fields and, therefore, there will be correct number of $2+2+1$ propagating degrees of freedom.

For parity-odd Lagrangian we find (again see Sec.~\ref{sec: Lagrangian construction for the first case} and Sec.~\ref{sec: Lagrangian construction for the second case} for the proof)
\begin{align}\label{L-po-EMS}
{L}^{po}_{\rm H} = g_1(\phi,X)\, F_{\alpha\beta}\tilde{F}^{\alpha\beta} \;,
\end{align}
where $g_1$ is an arbitrary function. Note that there is no higher-derivative terms in the parity-odd sector.

Now, let us explicitly show that the models presented in the literature, which might look different, are indeed subsets of our general Lagrangian \eqref{L-pe-EMS}.

The first case is
\begin{align*}
 L^{pe}_{\rm H}\big{|}_{w_1=4}
 &= 4 F_{\alpha }{}^{\gamma } F^{\alpha \beta } \phi_{\beta\gamma } - 2 F_{\alpha \beta } F^{\alpha \beta } \Box\phi\,, 
\end{align*}
where $L^{pe}_{\rm H}\big{|}_{w_1=4}$ means $w_1 = 4$ and all other functions are set to zero in \eqref{L-pe-EMS}. This Lagrangian coincides with Eq.~(8) of Ref.~\cite{Deffayet:2010zh}.

The second case is
\begin{align*}
L_{w_2}\big{|}_{w_2=4} &= \epsilon^{\mu\nu\rho\sigma}
\epsilon^{\alpha\beta\gamma\delta}\,
\phi_{\mu} \phi_{\alpha}\,
F_{\nu\rho} F_{\beta\gamma}\,
\phi_{\delta\sigma}\,, 
\end{align*}
which corresponds to the curved space generalization of Eq.~(14) of Ref.~\cite{Deffayet:2010zh}. 

The third case is the following combination
\begin{align}\label{L-HO-DF}
\begin{split}
&-\frac{1}{4} g_0\big{|}_{g_0 = X G_{5 \phi}} F_{\alpha\beta} F^{\alpha\beta}
+ g_{2}\big{|}_{g_2 = -\tfrac{1}{2}G_{5 \phi}} F^{\gamma\alpha}F_{\gamma}{}^{\beta} \phi_{\alpha} \phi_{\beta} 
+ L_{w_1}\big{|}_{w_1=- \tfrac{1}{2} X G_{5 X}} 
+ L_{w_2}\big{|}_{w_2=- \tfrac{1}{4} G_{5 X}} 
 \\
&= \frac{1}{2} G_{5}{}(\phi , X) 
\big( F_{\alpha }{}^{\gamma } F^{\alpha \beta } \phi_{\beta\gamma } 
+ F^{\alpha \beta } F_{\alpha \gamma }{}^{\gamma } \phi_{\beta }
- \tfrac{1}{4} F_{\alpha \beta } F^{\alpha \beta } \Box\phi \big)
\\
&+ \frac{1}{4} G_{5X} 
\big( F^{\alpha \beta } F^{\gamma \delta } \phi_{\beta } \phi_{\alpha\gamma } \phi {}_{\delta } 
- F_{\alpha }{}^{\gamma } F^{\alpha \beta } \phi_{\beta } \phi_{\gamma } \Box\phi \big) 
+ \mbox{total derivatives} \,,
\end{split}
\end{align}
which restores the Lagrangian $L_{5A}$ of Ref.~\cite{Mironov:2024umy} for the case of the frozen dilaton field.
Note that the Lagrangian \eqref{L-HO-DF} has quite complicated form which includes $\nabla_\rho{F}_{\mu\nu}$. However, this is nothing but a combination of the two interaction terms $w_{1,2}$ and some lower-order terms. 

The above analysis explicitly show that the four higher-derivative terms in Lagrangian \eqref{L-pe-EMS} serves as a full basis for the first and second subclasses of the HOMES theories defined by conditions $\textbf{I-1}$ to $\textbf{I-4}$ and $\textbf{II-1}$ to $\textbf{II-3}$ in Sec.~\ref{sec: Introduction}, respectively. Thus, our HOMES theory includes many models that appear in the existing literature as special subsets.

\section{Lagrangian construction for the first subclass}
\label{sec: Lagrangian construction for the first case}

In this section, we prove that the linear HOMES Lagrangians \eqref{L-pe-EMS} and \eqref{L-po-EMS} are the most general one for the first subclass that is defined by conditions $\textbf{I-1}$ to $\textbf{I-4}$ in Sec.~\ref{sec: Introduction}.

The field content of the model are one scalar field $\phi$, one vector field $A_\mu$, and metric $g_{\mu\nu}$. We need to find building blocks up to the second derivatives of $\{\phi,A_\mu,g_{\mu\nu}\}$. To zeroth-order in derivatives, we can consider any function of $\phi$ and cosmological constant in the scalar field and gravitational sectors respectively, while $A_\mu$ cannot be used due to the assumption of $U(1)$ gauge invariance. 

To the first-order in derivatives of the fields, we have the following building blocks~\cite{Fleury:2014qfa}
\begin{align}\label{eq: variables FXYZ}
X \equiv -\frac{1}{2} g^{\mu\nu} \phi_{\mu} \phi_{\nu} ,
\qquad
F \equiv -\frac{1}{4}F^{\mu\nu}F_{\mu\nu},
\qquad 
Y \equiv F_{\mu\nu}\tilde{F}^{\mu\nu}, 
\qquad
Z \equiv F^{\rho\mu}F_{\rho}{}^{\nu} \phi_{\mu} \phi_{\nu} \;,
\end{align}
where
\begin{align*}
F_{\mu\nu} \equiv \nabla_{\mu} A_{\nu} -  \nabla_{\nu} A_{\mu},
\qquad 
\tilde{F}^{\mu\nu} \equiv \frac{1}{2}\epsilon^{\mu\nu\rho\sigma}F_{\rho\sigma}\;,
\end{align*}
are the field strength and its Hodge dual which are totally antisymmetric tensors. Considering only terms that include up to quadratic order in the vector field and up to first order in the derivatives of all fields, the Lagrangian density takes the form
\begin{align}\label{L1-EMS}
L_{\rm MES} =  G_2(\phi, X) + g_{0}(\phi, X) F + g_{1}(\phi, X) Y + g_{2}(\phi,X) Z\;,
\end{align}
where $G_2$ and $g_i$ are arbitrary functions. Note that the zeroth-order terms, potential for the scalar field and cosmological constant are included in function $G_2$. Moreover, $g_{1}$ term breaks the parity.

Finding relevant terms with second-order derivatives is much more cumbersome. First of all, there are three different covariant terms, which include up to the linear order in the second derivatives, namely 
\begin{align}\label{fields-2ndD}
\{ \phi_{\mu\nu},\; \nabla_\rho{F}_{\mu\nu},\;R_{\mu\nu\sigma\rho} \} \;.
\end{align}
We should look at all possible contractions of \eqref{fields-2ndD} with metric $g^{\mu\nu}$, Levi-Civita tensor $\epsilon^{\mu\nu\rho\sigma}$, first derivative of the scalar field $\phi^\mu$, field strength of the vector field $F^{\mu\nu}$ that include up to quadratic order in vector field $A_\mu$. Thus, in general, the higher-order Lagrangian may take the following form
\begin{align}\label{L-HO-MES}
L_{\rm MES}^{\rm HO} = M^{\mu\nu} \phi_{\mu\nu} + M^{\mu\nu\rho} \nabla_\rho{F}_{\mu\nu} + M^{\mu\nu\sigma\rho} R_{\mu\nu\sigma\rho} \,.
\end{align} 
Our task is, first, to find the general forms of $M^{\mu\nu}$, $M^{\mu\nu\rho}$, and $M^{\mu\nu\sigma\rho}$, which are independent building blocks of the theory, and, second, to impose conditions that ensure second-order equations of motion, thereby simplifying the forms of $M^{\mu\nu}$, $M^{\mu\nu\rho}$, and $M^{\mu\nu\sigma\rho}$. We will do this procedure step-by-step in the next subsections.

\subsection{Independent building blocks of the action}\label{subsec: Independent building blocks in the action}

For the practical purposes, we first classify \eqref{L-HO-MES} as follows
\begin{align*}
L^{\rm HO}_{\rm MES} = L_S + L_{RC} + L_{PMN} + L_F + L_{RM} \,.
\end{align*}
Below, we will find the explicit forms of each Lagrangian in the expression above from which we can easily find the forms of $M^{\mu\nu}$, $M^{\mu\nu\rho}$, and $M^{\mu\nu\sigma\rho}$.

\subsection*{\boldmath${L}_S$} 
The Lagrangian density $L_S$ represents the terms which has the form of a scalar quantity $Q_i$ with no higher-derivative terms times another scalar quantity which contains one field and it is linear in second derivatives. For the latter scalar quantity, there exists only two scalar terms 
\begin{align*}
R,\;
\qquad
\Box\phi \,,
\end{align*}
where $R$ is the Ricci scalar. Taking into account the fact that the Lagrangian should include the terms which include up to quadratic order in vector field, we find
\begin{align*}
L_S = L_S^{pe} + L_S^{po},
\end{align*}
where $^{pe}$ and $^{po}$ denote parity-even and parity-odd parts, respectively and
\begin{align}\label{LS-pe}
L_S^{pe} &= 
(f_{0} F{} + f_{1} Z) \Box\phi + (f_{2} F + f_{3} Z) R,
\\ \nonumber
L_S^{po} &=  Y ( p_{0} \Box\phi + p_{1} R ) \,.
\end{align}
In this subsection, all $f_i$ and $p_i$ are functions of $\phi$ and $X$, but for simplicity, their dependence is not shown explicitly. Note that we do not include terms such as $s_{1}(\phi, X) \Box \phi$ or $s_{2}(\phi, X) R$, since these terms involve only the scalar field and cannot cancel contributions from the mixed terms. The pure scalar sector has already been fully explored, and its result is the KGB model; therefore, any term involving only the scalar field inevitably reduces to the KGB Lagrangian.

\subsection*{\boldmath$L_{RC}$ and \boldmath$L_{PMN}$} 
The Lagrangian for these terms have the forms: some tensor combination $M_{i}^{\mu\nu}$, which is constructed from $F_{\mu\nu}$, $\phi_\mu$, $\epsilon^{\mu\nu\rho\lambda}$, and $g^{\mu\nu}$ contracted with a term with two indices that contains only one field and is linear in the second derivatives. For the latter, there are only two options
\begin{align*}
R_{\mu\nu},\;
\qquad 
\phi_{\mu\nu}\;,
\end{align*}
where $R_{\mu\nu}$ is the Ricci tensor. Alternatively, one could use the traceless parts of $R_{\mu\nu}$ and $\phi_{\mu\nu}$ but the use of these only shifts the coefficients $f_i$ in \eqref{LS-pe}. In the following we shall adopt those with the traceless parts included. The Lagrangian $L_{RC}$ has the form $M^{\mu\nu}_{RC} R_{\mu\nu}$, while $L_{PMN}$ has the form $M^{\mu\nu}_{PMN}\phi_{\mu\nu}$ imposing $M_i^{\mu\nu}$ to be symmetric tensors. Before going further, it is convenient to introduce the auxiliary tensors $B^\mu_I$ as follows
\begin{align}\label{B-def}
B_0^\mu = \phi^{\mu} ,
\qquad
B_1^\mu =F_{\alpha }{}^{\mu } \phi^{\alpha } ,
\qquad
\tilde{B}_1^{\mu} = \tilde{F}_{\alpha}{}^{\mu } \phi^{\alpha} ,
\qquad
B_2^\mu =  F_{\alpha }{}^{\mu} F_{\beta }{}^{\alpha } \phi^{\beta} .
\end{align}
Note that $\tilde{B}_1^\mu$ cannot be confused with the Hodge dual of ${B}_1^{\mu}$ which has three free indices. One may notice that $B_1^\mu$ and $\tilde{B}_1^\mu$ characterize electric and magnetic parts of $F_{\mu\nu}$ in the case that $\phi^{\mu}$ is time-like. We do not need to consider terms like $F_{\alpha }{}^{\mu } \tilde{F}_{\beta }{}^{\alpha } \phi^{\beta}$ and $\tilde{F}_{\alpha }{}^{\mu } \tilde{F}_{\beta }{}^{\alpha } \phi^{\beta}$ since $\tilde{F}_{\mu \alpha } \tilde{F}_{\nu }{}^{\alpha }$ and $F_{\mu \alpha } \tilde{F}_{\nu }{}^{\alpha }$ could be expressed in terms of $F$, $Y$ and $F_{\mu \alpha } F_{\nu }{}^{\alpha }$ through the following identities:
\begin{align}
\label{eq: identities}
\begin{split}
F_{\mu \alpha } F_{\nu }{}^{\alpha } -  \tilde{F}_{\mu \alpha } \tilde{F}_{\nu }{}^{\alpha } &= 
\big( \nabla_{\beta }A_{\alpha } \nabla^{\beta }A^{\alpha } 
- \nabla_{\alpha }A_{\beta } \nabla^{\beta }A^{\alpha } \big) g_{\mu \nu }
= -2 F{} g_{\mu \nu }\;,
\\
F_{\mu \alpha } \tilde{F}_{\nu }{}^{\alpha } 
&= 
\epsilon_{\nu \rho \alpha \beta } 
\big(
\nabla^{\beta }A^{\alpha } \nabla^{\rho }A_{\mu} 
- \nabla^{\beta }A^{\alpha } \nabla_{\mu }A^{\rho}
\big) 
= \tfrac{1}{4} Y g_{\mu \nu } \;.
\end{split}
\end{align}

Now it is straightforward, to write down the Lagrangians. For parity-even sector, we find
\begin{align}\nonumber
L^{pe}_{RC} &= 
\bigl( f_{5} B_{0}^{\alpha } B_{1}^{\beta }
+ f_{6} B_{1}^{\alpha } B_{1}^{\beta }
+ f_{7} B_{0}^{\alpha } B_{2}^{\beta } 
+  f_{8} F^{\alpha \rho } F_{\rho }{}^{\beta}
+ f_{9} F  B_{0}^{\alpha } B_{0}^{\beta } 
+ f_{10} ZB_{0}^{\alpha } B_{0}^{\beta } \bigr) R_{\alpha \beta } \,,
\\
\label{L-PMN}
L^{pe}_{PMN} 
&= \bigl(
f_{12} B_{0}^{\alpha } B_{1}^{\beta }
+ f_{13} B_{1}^{\alpha } B_{1}^{\beta }
+ f_{14} B_{0}^{\alpha } B_{2}^{\beta }
+ f_{15} F^{\alpha \rho } F_{\rho }{}^{\beta }
+ f_{16} F B_{0}^{\alpha } B_{0}^{\beta }
+ f_{17} Z B_{0}^{\alpha } B_{0}^{\beta }
\bigr) \phi_{\alpha\beta} \,,
\end{align}
while for parity-odd sector, we find
\begin{align*}
L^{po}_{RC} 
&= \bigl(
p_{2} B_{0}^{\alpha } \tilde{B}_{1}^{\beta } 
+ p_{3} Y  B_{0}^{\alpha } B_{0}^{\beta } 
+ p_{4} B_{1}^{\alpha } \tilde{B}_{1}^{\beta }
\bigr) R_{\alpha \beta } \,,
\\
L^{po}_{PMN} 
&= \bigl(
p_{5} B_{0}^{\alpha } \tilde{B}_{1}^{\beta }
+ p_{6} Y B_{0}^{\alpha } B_{0}^{\beta }
+ p_{7} B_{1}^{\alpha } \tilde{B}_{1}^{\beta } \bigr) \phi_{\alpha\beta} \;. 
\end{align*}
By the same logic as before we do not include terms from the pure scalar sector, like: $s_{3}(\phi, X) B_{0}^{\alpha } B_{0}^{\beta } \phi_{\alpha\beta}$ and $s_4(\phi, X) B_{0}^{\alpha } B_{0}^{\beta } R_{\alpha\beta}$. Moreover, we do not consider the terms $\tilde{B}_1^{\alpha} \tilde{B}_1^{\beta} R_{\alpha\beta}$ and $\tilde{B}_1^{\alpha} \tilde{B}_1^{\beta} \phi_{\alpha\beta}$ in parity-even sector since they are not independent:
\begin{align*}
\tilde{B}_1^{\alpha} \tilde{B}_1^{\beta} R_{\alpha\beta} 
&= 
( Z - 4 F X ) R 
-2 X F_{\beta }{}^{\delta } F^{\beta \gamma } R_{\gamma \delta } 
+ \big( 2 F_{\alpha }{}^{\gamma } F_{\gamma }{}^{\delta } R_{\beta \delta }
-  F_{\alpha }{}^{\gamma } F_{\beta }{}^{\delta } R_{\gamma \delta } 
- 2 F{} R_{\alpha \beta } \big) \phi^\alpha \phi^\beta \;,
\\
\tilde{B}_1^{\alpha} \tilde{B}_1^{\beta} \phi_{\alpha\beta} &= 
(Z - 4 F X)\Box\phi 
- 2 X F_{\beta }{}^{\delta } F^{\beta \gamma } \phi_{\gamma\delta} 
+ \big( 
2 F_{\alpha }{}^{\gamma }  F_{\gamma }{}^{\delta } \phi_{\beta\delta}
- F_{\alpha }{}^{\gamma } F_{\beta }{}^{\delta } \phi_{\gamma\delta}
-2 F{}\phi_{\alpha\beta} \big) 
\phi^{\alpha }\phi^{\beta}
\;.
\end{align*}

\subsection*{\boldmath$L_{F}$} 
The Lagrangian $L_F$ has the form $\nabla^{\nu }F^{\eta \mu }$ times some combination of fields, which is linear in vector field and does not contain higher-order derivatives, i.e.: $M_{\nu\eta\mu}\nabla^{\nu }F^{\eta \mu }$. Let us discuss the structure of $M_{\nu\eta\mu}$. There are two distinct realizations of $M_{\nu\eta\mu}$. The first one, when all three indices come from the Levi-Civita tensor, i.e.: $B_i^{\rho}\epsilon_{\rho\nu\eta \mu}\nabla^{\nu }F^{\eta \mu }$. However this kind of terms are zero due to the Bianchi identity
\begin{align}
\nabla_{\alpha }F_{\mu \nu } + \nabla_{\mu }F_{\nu \alpha } + \nabla_{\nu }F_{\alpha \mu } = A^{\beta } (R_{\alpha \beta \mu \nu } + R_{\alpha \mu \nu \beta } -  R_{\alpha \nu \mu \beta }) = 0\;.
\end{align}
The second realization of $M_{\nu\eta\mu}\nabla^{\nu }F^{\eta \mu }$
is the following:  
\begin{align*}
M_{\nu\eta\mu}\nabla^{\nu }F^{\eta \mu }  = U_{j\;\mu\nu}B_{i\;\eta}\nabla^{\nu }F^{\eta \mu } .
\end{align*}
Interestingly, when $U_{j\;\mu\nu}$ is symmetric or antisymmetric with respect to $\mu$ and $\nu$, we do not need to consider other index orderings in the expression above due to the following identity:
\begin{align*} U_{j\;\mu\nu}B_{i\;\eta}\nabla^{\nu }F^{\eta \mu } &=  - U_{j\;\nu\eta}B_{i\;\mu}\nabla^{\nu }F^{\eta \mu } - U_{j\;\eta\mu}B_{i\;\nu}\nabla^{\nu }F^{\eta \mu } =  
U_{j\;\nu\mu}B_{i\;\eta}\nabla^{\nu }F^{\eta \mu } 
- U_{j\;\eta\mu}B_{i\;\nu}\nabla^{\nu }F^{\eta \mu }\;.
\end{align*}
Since the tensor $U_{j\;\mu\nu}$ can be decomposed into its symmetric and antisymmetric parts, the above statement is valid in general. Accordingly, if one chooses the tensor $U_{j\;\mu\nu}$ to be neither symmetric nor antisymmetric, then one should include two independent terms in the Lagrangian, namely $f_{A}(\phi,X)U_{j\;[\mu\nu]}B_{\eta}\nabla^{\nu }F^{\eta \mu }$ and $f_{S}(\phi,X)U_{j\;(\mu\nu)}B_{\eta}\nabla^{\nu }F^{\eta \mu }$, respectively. However, in practice, it is convenient to consider a different but equivalent basis, i.e.:
 $f_{n}(\phi,X)U_{j\;\mu\nu}B_{\eta}\nabla^{\nu }F^{\eta \mu }$ and $f_{r}(\phi,X)U_{j\;\nu\mu}B_{\eta}\nabla^{\nu }F^{\eta \mu }$.
We adopt this particular choice further in the text.

Based on the above points, the desired Lagrangian should have the following forms
\begin{align}\label{L-F-pe}
L_F^{pe} &= \nabla^{\nu }F^{\eta \mu }
\big[
\bigl( 
\tilde{f}_{18} B_{0}{}_{\eta }
+ \tilde{f}_{19} B_{1}{}_{\eta } \bigr)  g_{\mu \nu } 
+ \tilde{f}_{20} B_{0}{}_{\eta } F_{\mu \nu }
+ \tilde{f}_{21} B_{0}{}_{\mu } B_{0}{}_{\nu } B_{1}{}_{\eta }
\big]\;,
\\ \nonumber
L_F^{po} &= \nabla^{\nu }F^{\eta \mu }
\big(
p_{8} g_{\mu \nu } \tilde{B}_{1}{}_{\eta } 
+ p_{9} B_{0}{}_{\eta } \tilde{F}_{\mu \nu } 
+ p_{10} B_{0}{}_{\mu } B_{0}{}_{\nu } \tilde{B}_{1}{}_{\eta }
\big) \;.
\end{align}
Here, we do not write terms like $B_{0}{}_{\eta } B_{0}{}_{\nu } \tilde{B}_{1}{}_{\mu }$ and  $B_{0}{}_{\eta } B_{0}{}_{\nu } B_{1}{}_{\mu }$, since they are redundant due to the identity
\begin{align*}
B_{i}{}_{\mu } B_{i}{}_{\nu } B_{j}{}_{\eta } \nabla^{\nu }F^{\eta \mu } + B_{i}{}_{\eta } B_{i}{}_{\nu } B_{j}{}_{\mu } \nabla^{\nu }F^{\eta \mu } = 0\;,
\end{align*}
and already included in $p_8$ and $f_{21}$ terms, respectively. Moreover, we do not write term
$\epsilon_{\mu \nu }{}^{\chi \rho } \tilde{F}_{\delta \chi } \phi^{\delta} \phi_{\eta} \phi_{\rho} \nabla^{\nu }F^{\eta \mu }$ because it can be expressed in terms of $\tilde{f}_{20}$ and $\tilde{f}_{21}$:
\begin{align*}
\epsilon_{\mu \nu }{}^{\chi \rho } \tilde{F}_{\delta \chi } \phi^{\delta} \phi_{\eta} \phi_{\rho} \nabla^{\nu }F^{\eta \mu } 
&=
-2X F^{\gamma \delta } \nabla_{\delta }F_{\beta \gamma } \phi^{\beta}
+ F_{\alpha }{}^{\delta } \phi^{\alpha} \phi^{\beta} \nabla_{\gamma }F_{\beta \delta } \phi^{\gamma} 
\\
&=
\big( \tilde{f}_{20}\big|_{\tilde{f}_{20} \to -2X} F_{\mu \nu } \phi_{\eta} 
+ \tilde{f}_{21}\big|_{\tilde{f}_{21} \to -1} F_{\sigma \eta } \phi_{\mu} \phi_{\nu} \phi^{\sigma} 
\big) \nabla^{\nu }F^{\eta \mu }  .
\end{align*}

Before going to the next Lagrangian, let us further simplify \eqref{L-F-pe}. Doing appropriate integration by parts and using the identity $F^{\mu\alpha}\nabla_\alpha{F}_{\eta\mu}=\nabla_\eta{F}=-\nabla_\eta(F_{\alpha\beta}F^{\alpha\beta})/4$, it is straightforward to show that
\begin{align}\label{L-F-pe-2}
L_F^{pe} &=
L_{\rm MES} + L_S^{pe} + L^{pe}_{PMN}
+ \mbox{total derivatives} \;,
\end{align}
with
\begin{align}\nonumber
\begin{split} 
g_0 &= - 2X \partial_\phi (\tilde{f}_{19}-\tilde{f}_{20}) \,,
\qquad
g_{2} = \tilde{f}_{19,\phi} - X \tilde{f}_{21,\phi} \,,
\qquad
G_2 = g_1 = 0 \,,
\\
f_0 &= \tilde{f}_{19}-\tilde{f}_{20} \,,
\qquad
f_{1} = \frac{1}{2} \tilde{f}_{21} \,,
\qquad
f_{2,3} = 0 \,,
\qquad
f_{12} = \tilde{f}_{18,X} \,,
\qquad
f_{14} = - (\tilde{f}_{19,X} - \tilde{f}_{21}) \,,
\\
f_{15} &= - \tilde{f}_{19} \,,
\qquad
f_{16} = -\partial_X (\tilde{f}_{19} - \tilde{f}_{20}) \,,
\qquad
f_{17} = - \frac{1}{2} \tilde{f}_{21,X} \,,
\qquad
f_{11,13} = 0 \,.
\end{split}
\end{align}
The above result explicitly shows that the Lagrangian $L_F^{pe}$, defined in \eqref{L-F-pe}, can be completely expressed in terms of the Lagrangians \eqref{L1-EMS}, \eqref{LS-pe}, and \eqref{L-PMN} and, therefore, it is redundant. This is interesting since it shows that, at least in the parity-even sector, even if we include $\nabla_\rho{F}_{\mu\nu}$ in the first subclass of HOMES, it can be always rewritten in terms of $\phi_{\mu\nu}$ and some lower-order terms.  

\subsection*{\boldmath$L_{RM}$} 
Finally, we turn to $L_{RM}$, which includes a contraction between the Riemann tensor and a four indices object $M^{\mu\nu\rho\eta}_{RM}$. There are two distinct realizations of $M^{\mu\nu\rho\eta}_{RM}$. In the first case, three indices are coming from the Levi-Civita tensor; however, in this case, the contraction with the Riemann tensor identically vanishes due to the first Bianchi identity for the Riemann tensor. In the second case, $M^{\mu\nu\rho\eta}_{RM}$ can be expressed as a product of two tensors
\begin{align*}
M^{\mu\nu\rho\eta}_{RM} = V_{i}^{\mu\nu}V_{j}^{\rho\eta}\;.
\end{align*}
Due to the symmetries $R_{\mu \nu \eta \rho } =- R_{\nu \mu \eta \rho }$, $R_{\eta \rho \mu \nu } = R_{\mu \nu \eta \rho }$, and the first Bianchi identity $R_{\mu \eta \rho \nu } + R_{\mu \nu \eta \rho } + R_{\mu \rho \nu \eta } = 0$, it is sufficient to consider only the following contraction
\begin{align*}
R_{\mu \nu \eta \rho } V_{i}{}^{\mu \rho } V_{j}{}^{\nu \eta }\;.
\end{align*}
In addition, if $V_{i}{}^{\mu \rho }$ is symmetric/antisymmetric, only the symmetric/antisymmetric part of $V_{j}{}^{\nu \eta }$ contributes. Finally, the order of $V_{i}{}^{\mu \rho }$ and $V_{j}{}^{\nu \eta }$ is irrelevant
\begin{align*}
R_{\mu \nu \eta \rho } V_{i}{}^{\mu \rho } V_{j}{}^{\nu \eta } = R_{\mu \nu \eta \rho } V_{j}{}^{\mu \rho } V_{i}{}^{\nu \eta }\;. 
\end{align*}
For the sake of presentation, it is convenient to introduce the following auxiliary tensors
\begin{align}\nonumber
\begin{split}
V_1^{\mu\rho} = F^{\mu\rho},
\qquad
V_{01}^{\mu\rho} = B_{0}^{\mu}B_{1}^{\rho},
\qquad
\tilde{V}_{10}^{\mu\rho} = B_{0}^{\mu}\tilde{B}_{1}^{\rho} \,.
\end{split}
\end{align}
Here we do not consider $B^{\mu}_{1} B^{\rho}_{1},\;\tilde{B}^{\mu}_{1} B^{\rho}_{1},\,\cdots $ because
\begin{align*}
R_{\mu \nu \eta \rho } B^{\mu}_{1} B^{\rho}_{1}
B^{\nu}_{0} B^{\eta}_{0}  &=  R_{ \nu \mu \rho \eta  } B^{\mu}_{1} B^{\rho}_{1}
B^{\nu}_{0} B^{\eta}_{0} =  R_{ \nu \mu \rho \eta  } V_{01}^{\eta\mu} V_{{01}}^{\nu\rho} \\
&= 
- R_{ \nu \mu \eta \rho   } V_{01}^{\eta\mu} V_{{01}}^{\nu\rho} = - R_{ \mu \nu \eta \rho   } V_{01}^{\eta\nu} V_{{01}}^{\mu\rho},\;\cdots \,.
\end{align*}
We can also use the Hodge dual of the auxiliary tensors $\epsilon^{\mu\rho}{}_{\alpha\beta}V^{\alpha\beta}_{i}$ as building blocks.\footnote{Note that, for the sake of simplicity, we omit the $1/2$ factor, since it corresponds to a redefinition of the coefficients $f_i$ and $p_i$ in the Lagrangians and thus does not affect the final results.} We do not need to consider the Hodge dual of $\tilde{V}_{10}^{\mu\rho}$ as it is not an independent building block $\epsilon^{\mu \rho }{}_{\alpha \beta } B_{0}^{\alpha } \tilde{B}_{1}^{\beta}= V_{01}^{\mu\rho} - V_{01}^{\rho\mu} + 2 X V_{1}^{\mu\rho}$. Moreover, we could consider another building block $V_{02}^{\mu\rho} = B_{0}^{\mu}B_{2}^{\rho}$, which is quadratic in the vector field, and can only comes together with $B_0^{\mu}B_0^{\nu}$ that is zeroth-order in vector field. However, all possible contractions of $V_{02}^{\mu\rho}$ and $B_0^{\mu}B_0^{\nu}$ with the Riemann tensor, like $R_{\mu \nu \eta \rho } V_{02}^{\mu\rho} V_{0}^{\nu\eta} = 0$ and $R_{\mu \nu \eta \rho } V_{02}^{\rho\mu} V_{0}^{\nu\eta} = 0$, vanish and we do not need to consider it. Also we do not consider $R_{\mu \nu \eta \rho } V_{{01}}^{\mu\rho} V_{{01}}^{\nu\eta} = 0, R_{\mu \nu \eta \rho } V_{{01}}^{\rho\mu} V_{{01}}^{\eta\nu} = 0,\,\cdots$, as these terms vanish. It is interesting to note that the possible building block $B_0^{\mu}B_0^{\rho}$ never appears in the Lagrangian. The reason for this is as follows. First, $B_0^{\mu}B_0^{\rho}$ is symmetric, which means the second block cannot be antisymmetric. Therefore, we are left with only three options for the second building block: $B_0^{\mu}B_0^{\rho}, V_{01}^{\mu\rho}$, and $\tilde{V}_{01}^{\mu\rho}$. However, all three of these contractions vanish identically due to the symmetries of the Riemann tensor. Finally, we do not consider the building block  $F^{\sigma\rho } F_{\sigma }{}^{\mu },$  since the only combination involving this block is  $R_{\mu \nu \eta \rho } F^{\sigma\rho } F_{\sigma }{}^{\mu } \phi^{\eta} \phi^{\nu} $. However, this term is not independent
\begin{align*}
&F^{\sigma\rho } F_{\sigma }{}^{\mu } R_{\mu \beta \alpha \rho } \phi^{\alpha} \phi^{\beta}  = 
f_{7}\big|_{f_{7} = -1} F_{\alpha }{}^{\gamma } F_{\gamma }{}^{\delta } R_{\beta \delta} \phi {}^{\alpha } \phi {}^{\beta } 
+ f_{9}\big|_{f_{9} = 2} F R_{\alpha \beta } \phi {}^{\alpha } \phi {}^{\beta }
\\
&
- f_{19}\big|_{f_{19} = 1}
 F_{\alpha }{}^{\gamma } F^{\delta \eta } R_{\beta \delta \gamma \eta } \phi {}^{\alpha } \phi {}^{\beta }
- \tfrac{1}{2} f_{21}\big|_{f_{21} = 1} 
  \epsilon_{\gamma \delta \mu \nu } F^{\gamma \delta } R_{\beta }{}^{\mu }{}_{\eta }{}^{\nu } \tilde{F}_{\alpha }{}^{\eta } \phi {}^{\alpha } \phi {}^{\beta }
\;.
\end{align*}
Therefore, we only need to use $\{V_1^{\mu\rho}, \tilde{V}_1^{\mu\rho}, {V}_{01}^{\mu\rho}, \tilde{V}_{01}^{\mu\rho}, \tilde{V}_{10}^{\mu\rho}\}$. Taking into account all of these symmetries and identities, we find
\begin{align*}
L^{pe}_{RM} 
&= R_{\mu \nu \eta \rho }
\big(
f_{18} V_{1}^{\mu \rho } V_{1}^{\nu \eta } 
+ f_{19} V_{1}^{\mu \rho } V_{{01}}^{\nu \eta }  
+ f_{20} \tilde{V}_{1}^{\mu \rho } \tilde{V}_{1}^{\nu \eta } 
+ f_{21} \tilde{V}_{1}^{\mu \rho } \tilde{V}_{10}^{\nu \eta } 
+ f_{22} \tilde{V}_{1}^{\mu \rho } \tilde{V}_{{01}}^{\nu \eta }  
+ f_{23} \tilde{V}_{10}^{\mu \rho } \tilde{V}_{{01}}^{\nu \eta } 
\\
&\hspace{1.7cm}
+ f_{24} \tilde{V}_{{01}}^{\mu \rho } \tilde{V}_{{01}}^{\nu \eta } 
+ f_{25} V_{{01}}^{\rho \mu } V_{{01}}^{\nu \eta } 
+ f_{26} \tilde{V}_{10}^{\rho \mu } \tilde{V}_{10}^{\nu \eta }
\big)\;,
\\
L^{po}_{RM} &= R_{\mu \nu \eta \rho }
\big(
p_{11} V_{1}^{\mu \rho } \tilde{V}_{1}^{\nu \eta } 
+ p_{12} V_{1}^{\mu \rho } \tilde{V}_{10}^{\nu \eta } 
+ p_{13} V_{1}^{\mu \rho }  \tilde{V}_{{01}}^{\nu \eta }
+ p_{14} \tilde{V}_{1}^{\mu \rho } V_{{01}}^{\nu \eta }  
+ p_{15} V_{{01}}^{\mu \rho } \tilde{V}_{{01}}^{\nu \eta }
+ p_{16} V_{{01}}^{\rho \mu } \tilde{V}_{10}^{\nu \eta }
\big)\;.
\end{align*}
Moreover, not all of the above combinations are independent. Using the \textit{xAct} package for \textit{Mathematica}, one can prove that some terms above can be expressed through each other. We define ${\cal T}_i$ as the coefficient of $f_i$, for example ${\cal T}_0$ is $F\Box{\phi}$. Then it is straightforward to show that ${\cal T}_{23}= - X {\cal T}_{22}- {\cal T}_{24}$ and ${\cal T}_{26}= 2 X {\cal T}_{21} - 2 X {\cal T}_{22} - 2 {\cal T}_{24}$ which shows that, without loss of generality, we can set $f_{23}$ and $f_{26}$ to zero. After that we sum all pieces together and obtain parity-preserving and parity-violating Lagrangians. However, if one sums all Lagrangians $L_S,\;L_{RC},\;\ldots$ together, then again not all terms are independent. Similarly, one has ${\cal T}_{18} = -2 {\cal T}_{2} + 2 {\cal T}_{8} - {\cal T}_{20}$, ${\cal T}_{22} =  \tfrac{1}{2} {\cal T}_{3} - {\cal T}_{6} + {\cal T}_{7} -  {\cal T}_{19}$, ${\cal T}_{24} = - \tfrac{X}{2} {\cal T}_{3} + X {\cal T}_{6} -\tfrac{1}{2} {\cal T}_{10}+ \tfrac{1}{2} {\cal T}_{25}$. Thus, without loss of generality, we can also set the functions $f_{18},\;f_{22},$ $f_{24}$ to zero.

Collecting all terms one arrives at the general parity-even Lagrangian
\begin{align}\label{eq: pe starting action}
\begin{split}
{L}^{{\rm HO}, pe}_{\rm MES} 
&= f_{0} F \Box\phi
+ f_{1} Z \Box\phi 
+ f_{2} F R 
+ f_{3} R Z 
+ f_{5} F_{\alpha }{}^{\gamma } R_{\beta \gamma } \phi {}^{\alpha } \phi {}^{\beta } 
+ f_{6} F_{\alpha }{}^{\gamma } F_{\beta }{}^{\delta } R_{\gamma \delta } \phi {}^{\alpha } \phi {}^{\beta } 
\\
&+ f_{7} F_{\alpha }{}^{\gamma } F_{\gamma }{}^{\delta } R_{\beta \delta } \phi {}^{\alpha } \phi {}^{\beta } 
- f_{8} F_{\alpha }{}^{\gamma } F^{\alpha \beta } R_{\beta \gamma }
+ \left( f_{9} F + f_{10} Z \right) R_{\alpha \beta } \phi {}^{\alpha } \phi {}^{\beta } 
+ f_{12} F_{\alpha }{}^{\gamma }  \phi {}^{\alpha } \phi {}^{\beta } \phi {}_{\beta }{}_{\gamma } 
\\
&+ \left( f_{13} F_{\alpha }{}^{\gamma } F_{\beta }{}^{\delta } \phi {}_{\gamma }{}_{\delta } 
+ f_{14} F_{\alpha }{}^{\gamma } F_{\gamma }{}^{\delta }\phi {}_{\beta }{}_{\delta } \right)  \phi {}^{\alpha } \phi {}^{\beta } 
-  f_{15} F_{\alpha }{}^{\gamma } F^{\alpha \beta } \phi {}_{\beta}{}_{\gamma } 
+ \left( f_{16} F + f_{17} Z \right) \phi {}^{\alpha } \phi {}_{\alpha }{}_{\beta } \phi {}^{\beta } 
\\
&-  f_{19} F_{\alpha }{}^{\gamma } F^{\delta \zeta } R_{\beta \delta \gamma \zeta } \phi {}^{\alpha } \phi {}^{\beta } 
-f_{20} (2F R +2 F_{\alpha }{}^{\gamma } F^{\alpha \beta } R_{\beta \gamma } - F^{\alpha \beta } F^{\gamma \delta } R_{\alpha \gamma \beta \delta }) 
\\
&- f_{21} \tilde{F}^{\eta \mu} \tilde{F}_{\alpha }{}^{\zeta } R_{\beta\eta\zeta\mu} \phi {}^{\alpha } \phi {}^{\beta }
+ f_{25} F_{\alpha }{}^{\zeta } F_{\beta }{}^{\eta } R_{\gamma \zeta \delta \eta } \phi {}^{\gamma } \phi {}^{\delta } \phi^{\alpha } \phi^{\beta } \;, 
\end{split}
\end{align}
and parity-odd Lagrangian
\begin{align}
\label{eq: po starting action}
\begin{split}
{L}^{{\rm HO}, po}_{\rm MES} 
&= p_{0} Y \Box\phi  
+ p_{1} Y R
+ p_{2} R_{\alpha }{}^{\gamma } \tilde{F}_{\beta \gamma } \phi {}^{\alpha } \phi {}^{\beta }
+ p_{3} Y R_{\alpha \beta } \phi {}^{\alpha } \phi {}^{\beta } 
+ p_{4} F_{\alpha }{}^{\gamma } R_{\gamma }{}^{\delta } \tilde{F}_{\beta \delta } \phi {}^{\alpha } \phi {}^{\beta } 
\\
&+ p_{5} \tilde{F}_{\alpha }{}^{\gamma } \phi {}^{\alpha } \phi {}^{\beta } \phi {}_{\beta }{}_{\gamma } 
+ p_{6} Y \phi {}^{\alpha } \phi {}_{\alpha }{}_{\beta } \phi {}^{\beta } 
+ p_{7} F_{\alpha }{}^{\gamma }  \tilde{F}_{\beta }{}^{\delta } \phi {}^{\alpha } \phi {}^{\beta } \phi {}_{\gamma }{}_{\delta } 
+ p_{8} \tilde{F}_{\alpha }{}^{\beta } \phi {}^{\alpha } \nabla_\gamma F_{\beta }{}^{\gamma }
\\
&+ p_{9} \tilde{F}^{\beta \gamma } \phi {}^{\alpha } \nabla_\gamma F_{\alpha \beta }
- p_{10} \tilde{F}_{\alpha }{}^{\delta } \phi {}^{\alpha } \phi {}^{\beta }  \phi^{\gamma }  \nabla_\gamma F_{\beta \delta } 
- \tfrac{1}{2} p_{11} \epsilon_{\gamma \delta \zeta \eta } F^{\alpha \beta } F^{\gamma \delta }  R_{\alpha }{}^{\zeta }{}_{\beta }{}^{\eta }
\\
&+ \left( p_{12} F^{\gamma \delta } R_{\beta \gamma \delta \zeta } \tilde{F}_{\alpha }{}^{\zeta }
- \tfrac{1}{2} p_{13} \epsilon_{\beta \gamma \eta \mu } F_{\alpha }{}^{\gamma } F^{\delta \zeta } R_{\delta }{}^{\eta }{}_{\zeta }{}^{\mu }
- \tfrac{1}{2} p_{14} \epsilon_{\delta \zeta \eta \mu } F_{\alpha }{}^{\gamma } F^{\delta \zeta }  R_{\beta }{}^{\eta }{}_{\gamma }{}^{\mu }  
\right) \phi {}^{\alpha } \phi {}^{\beta }
\\
 &- \left( \tfrac{1}{2} p_{15} \epsilon_{\delta \eta \mu \nu } F_{\alpha }{}^{\zeta } F_{\beta }{}^{\eta } R_{\gamma }{}^{\mu }{}_{\zeta }{}^{\nu }
 - p_{16} F_{\alpha }{}^{\zeta } R_{\gamma \zeta \delta \eta } \tilde{F}_{\beta }{}^{\eta } 
 \right)
 \phi {}^{\alpha } \phi {}^{\beta } \phi {}^{\gamma } \phi {}^{\delta } 
\;,
\end{split}
\end{align}
where $f_i$ and $p_i$ are functions of $\phi$ and $X$. We omit the pure scalar sector in the parity-even part, as it is well known that this contribution leads only to the KGB Lagrangian. Including the pure scalar terms 
\begin{align}\label{scalar-couplings}
s_{1}(\phi, X)\Box \phi \,, 
\qquad
s_{2}(\phi, X)R \,, 
\qquad
s_{3}(\phi, X)\phi^\alpha\phi^\beta\phi_{\alpha\beta} \,,
\qquad
s_{4}(\phi, X)\phi^\alpha\phi^\beta R_{\alpha\beta} \,,
\end{align}
in the parity-even sector leads to a total Lagrangian containing $(20+4)+17=41$ arbitrary functions of the two variables $\phi$ and $X$. The explicit forms of $M^{\mu\nu}$, $M^{\mu\nu\rho}$, and $M^{\mu\nu\sigma\rho}$ in \eqref{L-HO-MES} can be readily obtained from Eqs.~\eqref{eq: pe starting action}, \eqref{eq: po starting action}, and \eqref{scalar-couplings}.

\subsection{Imposing second-order equations of motion}\label{subsec: Imposing second-order equations of motion}

Having found all independent terms in the Lagrangians \eqref{eq: pe starting action} and \eqref{eq: po starting action}, we impose the condition that equations of motion of all fields $\{\phi,A^\mu,g^{\mu\nu}\}$ should be second-order. Indeed, this is not an easy task in practice if we start from the covariant equations of motion. To simplify the task, we demand second-order equations of motion for the scalar and vector fields for specific classes of configurations including some arbitrary functions
\begin{align}
\label{eq: field configurations}
\begin{split}
&g^{(1,2)}_{\mu\nu} = \text{diag}\big( -1,a^2_1(t,x),a^2_2(t,x),a^2_3(t,x) \big),
\\ 
&A^{(1)}_{\mu} = \big( 0, 0, A_2(t,x), 0 \big), 
\hspace{2.7cm}
\phi^{(1)} = \phi(t),\; 
\\ 
&A^{(2)}_{\mu} = \big( 0, A_1(t,x), A_2(t,x), A_3(t,x) \big), 
\quad
\phi^{(2)} = \phi(t,x) \,.
\end{split}   
\end{align}
This gives us {\it necessary conditions}. We then substitute back these conditions into the covariant Lagrangian. We repeat this strategy with different background configurations until, after substitution, the covariant equations of motion for all fields become second-order indicating that we have found the {\it sufficient conditions} which ensure second-order equations of motion for all fields.

The above process will impose relations within the $f_i(\phi,X)$ and within the $p_i(\phi,X)$, but not between them. We thus study the parity-even and parity-odd sectors separately in the following subsections.

\subsubsection{Parity-even sector}
\label{subsubsec: The parity-even part}

Taking variation of the parity-even Lagrangian \eqref{eq: pe starting action} with respect to $\phi$ and $A^\mu$, and then substituting \eqref{eq: field configurations} and demanding that the equations of motion be second-order, we find some (differential) relations between $f_i$. After imposing these relations, the Lagrangian \eqref{eq: pe starting action} simplifies to
\begin{align}
\label{L-nc}
\begin{split}
{L}^{{\rm HO}, pe}_{\rm MES}\big|_{\text{n.c.}} 
=
L_{KGB}
+ L_{w_0} + L_{f_0} + L_{f_1} + L_{f_{21}} \,,
\end{split}
\end{align}
where the subscript $\text{n.c.}$ denotes the necessary conditions and we have defined
\begin{align}
\label{L-f_i-def}
\begin{split}
 L_{KGB} &\equiv G_3(\phi,X) \Box \phi\,,
\hspace{2.3cm}
L_{w_0} \equiv 
w_{0}{}(\phi) R_{\alpha \gamma \beta \delta } \tilde{F}^{\alpha \beta } \tilde{F}^{\gamma \delta }  \,,
\\
L_{f_0} &\equiv 
\tfrac{1}{2} f_{0}{}(\phi , X) \tilde{F}_{\alpha }{}^{\gamma } \tilde{F}^{\alpha \beta } \phi_{\beta }{}_{\gamma }\;,
\qquad
L_{f_1} \equiv 
f_{1}{}(\phi , X) \big( 
2 \tilde{F}_{\alpha }{}^{\gamma } \tilde{F}^{\alpha \beta } X \phi {}_{\beta }{}_{\gamma } 
+ \tilde{F}_{\alpha }{}^{\gamma } \tilde{F}_{\beta }{}^{\delta } \phi {}^{\alpha } \phi {}^{\beta } \phi {}_{\gamma}{}_{\delta } 
\big) \;,
\\
L_{f_{21}} &\equiv f_{21}{}(\phi , X) 
\Big[
\tfrac{1}{3} X F^{\beta \gamma } F^{\delta \zeta } (R_{\beta \gamma \delta \zeta } + R_{\beta \delta \gamma \zeta })
-2X F_{\beta }{}^{\delta } F^{\beta \gamma } R_{\gamma \delta }
-2X F R
 \\
 &+ \big( 
 2 F_{\alpha }{}^{\gamma } F_{\gamma }{}^{\delta } R_{\beta \delta }
 -2F R_{\alpha \beta }
 -  F_{\alpha }{}^{\gamma } F_{\beta }{}^{\delta } R_{\gamma \delta } 
 + \tfrac{1}{2} F_{\alpha }{}^{\gamma } F_{\beta \gamma } R  
- F_{\gamma }{}^{\zeta } F^{\gamma \delta } R_{\alpha \delta \beta \zeta } 
\\
 &+ \tfrac{2}{3} F_{\alpha }{}^{\gamma } F^{\delta \zeta } R_{\beta \gamma \delta \zeta }
 + \tfrac{2}{3} F_{\alpha }{}^{\gamma } F^{\delta \zeta } R_{\beta \delta \gamma \zeta } \big) \phi^{\alpha } \phi^{\beta }
 \Big] \;.
\end{split}
\end{align}
We have added the KGB term in \eqref{L-nc} for completeness. Indeed, taking into account the four scalar terms \eqref{scalar-couplings} simply gives us one independent term, which is nothing but KGB. Note that our setup can be thought of as a natural generalization of KGB that includes a $U(1)$ vector field. Here 

While not obvious, the $f_{21}$ term vanishes identically due to the dimension-dependent identity that antisymmetrization over five indices in four dimensions is zero
\begin{align*}
L_{f_{21}} = -\tfrac{5!}{16} \, f_{21} \, g^{\alpha \gamma } g^{\beta \mu } g^{\nu \rho } g^{\xi \zeta } g^{\sigma \delta } \,
F_{\gamma \mu } F_{[\alpha \beta }   \phi_{\xi } R_{\nu \sigma] \rho \delta }  \phi {}_{\zeta } = 0 \;.
\end{align*}
Thus, without loss of generality, we can set it to zero
\begin{align}\label{f-zero-21}
\boxed{f_{21}(\phi,X) = 0 \,.}
\end{align}

In summary, from $20+4$ coupling functions in \eqref{eq: pe starting action} and \eqref{scalar-couplings} we ended up to 4 functions $w_{0},\;f_{0},\;f_{1},\;G_3$. Taking variation of the Lagrangian ${L}^{pe}\big|_{\text{n.c.}}$ with respect to scalar field, vector field, and metric, one can directly confirm that $w_0, f_0, f_1, G_{3}$ terms do not generate any higher-order derivatives in the equations of motion. Therefore, we have found necessary and sufficient conditions. 

The $w_0$ term corresponds to the Horndeski non-minimal coupling \cite{Horndeski:1976gi} in the vector field sector while the term with $G_{3}$ in the scalar sector is the KGB term. The $f_{0,1}$ are potentially new terms which characterize higher-derivative interactions between scalar and vector sectors. We should be able to recover the $U(1)$ symmetric result of Ref. \cite{Heisenberg:2018acv} as a subset. Indeed, by only focusing on the second derivatives of the scalar field $\phi_{\mu\nu}$, two $U(1)$-invariant interactions between scalar and vector field sectors have been found in \cite{Heisenberg:2018acv}. We have proved in \eqref{L-F-pe-2}, all terms that include $\nabla_\rho{F}_{\mu\nu}$ can be rewritten in terms of $\phi_{\mu\nu}$ and some lower-order derivative terms. Moreover, considering all possible terms that include $\phi_{\mu\nu}$, we still have two interaction terms between scalar and vector field sectors. This means the Lagrangian that is found in Ref. \cite{Heisenberg:2018acv} is indeed the most general one that can generate second-order equations of motion as far as the Lagrangian is quadratic in $F_{\mu\nu}$. The map between our mixing terms $f_{0,1}$ and the two mixing terms that are found in Ref. \cite{Heisenberg:2018acv} becomes manifest by defining
\begin{align}
w_1 \equiv \frac{1}{2} f_0 + 2X f_1 \,,
\qquad
w_2 = f_1 \,.
\end{align}
Working with $w_{1,2}$ instead of $f_{0,1}$, and adding up the lower-order part \eqref{L1-EMS} to the above results, we deduce that, in the first subclass, the \textit{most} general Lagrangian which does not lead to higher-order derivatives in the all equations of motion reads:  
\begin{align}\label{L-Homes-1st-Class}
\begin{split}
L^{pe}_{\rm H} &= g_0 (\phi, X) F + g_2(\phi,X)  Z +  G_2 (\phi, X) - G_3(\phi,X)\,\square\phi 
\\
&+ \big[
w_0(\phi)\,R_{ \beta \delta \alpha \gamma}
+ (w_1(\phi,X) g_{\beta\delta} 
+ w_2(\phi,X) \phi_{\beta} \phi_{\delta}) \phi_{\alpha\gamma} 
\big] \tilde{F}^{\alpha \beta } \tilde{F}^{\gamma \delta } \;.
\end{split}
\end{align}
This completes the proof that, in the even-parity sector, the most general Lagrangian defined by conditions $\textbf{I-1}$ to $\textbf{I-4}$ in Sec.~\ref{sec: Introduction} is given by  \eqref{L-pe-EMS}.

\subsubsection{Parity-odd sector}\label{subsubsec: The parity-odd part}
Following the same reasoning as in the previous section, we will begin by obtaining the necessary conditions for the absence of higher-order derivatives in the equations of motion generated by the parity-odd Lagrangian \eqref{eq: po starting action}. We thus take the variation of \eqref{eq: po starting action} with respect to the scalar field and vector field, then substituting the field configurations \eqref{eq: field configurations} and require the absence of higher-order derivatives. These requirements lead to the set of necessary conditions between $p_i$ which after substituting back into the parity-odd Lagrangian \eqref{eq: po starting action} gives 
\begin{align}\label{L-nc-po}
&{L}^{{\rm HO}, po}_{\rm MES}\big|_{\text{n.c.}} = L_{p_3} + L_{p_5}  + L_{p_9} +  L_{p_{13}} + L_{p_{16}}\,,
\end{align}
where we have defined
\begin{align}
\label{L-p_i-def}
\begin{split}
L_{p_3} &\equiv  
p_{3}{}(\phi , X) 
(R X Y 
-2 \epsilon_{\gamma \delta \eta \mu } F^{\alpha \beta } F^{\gamma \delta } R_{\alpha }{}^{\eta }{}_{\beta }{}^{\mu } X 
- 2 \epsilon_{\delta \eta \mu \nu } F_{\alpha }{}^{\gamma } F^{\delta \eta } R_{\beta }{}^{\mu }{}_{\gamma }{}^{\nu } \phi {}^{\alpha } \phi {}^{\beta }
\\
&+ 4 F^{\gamma \delta } R_{\beta \gamma \delta \eta } \tilde{F}_{\alpha }{}^{\eta } \phi {}^{\alpha } \phi {}^{\beta } 
+ 4 F_{\alpha }{}^{\gamma } R_{\gamma }{}^{\delta } \tilde{F}_{\beta \delta } \phi {}^{\alpha } \phi {}^{\beta } + R_{\alpha \beta } Y \phi {}^{\alpha } \phi {}^{\beta })
\,,
\\
L_{p_5} &\equiv \tfrac{1}{2} p_{5}{}(\phi , X) \epsilon_{\beta \gamma \delta \zeta } F^{\gamma \delta }  \phi {}^{\alpha } \phi {}^{\beta } \phi {}_{\alpha }{}^{\zeta } \, ,
\\ 
L_{p_9} &\equiv  
p_{9}{}(\phi , X) 
(\tilde{F}^{\beta \gamma } \phi {}^{\alpha } \nabla_\gamma F_{\alpha \beta }
- \tfrac{1}{4} Y \square\phi) + \tfrac{1}{4}  p_{9 X} Y \phi {}^{\alpha } \phi {}_{\alpha }{}_{\beta } \phi {}^{\beta } 
\,,
\\
L_{p_{13}} &\equiv  
p_{13}{}(\phi , X) 
(
F^{\gamma \delta } R_{\beta \gamma \delta \eta } \tilde{F}_{\alpha }{}^{\eta } \phi {}^{\alpha } \phi {}^{\beta } - \tfrac{1}{2} \epsilon_{\gamma \delta \eta \mu } F^{\alpha \beta } F^{\gamma \delta } R_{\alpha }{}^{\eta }{}_{\beta }{}^{\mu } X -  \tfrac{1}{2} \epsilon_{\beta \gamma \mu \nu } F_{\alpha }{}^{\gamma } F^{\delta \eta } R_{\delta }{}^{\mu }{}_{\eta }{}^{\nu } \phi {}^{\alpha } \phi {}^{\beta } 
),
\\
L_{p_{16}} &\equiv  
p_{16}{}(\phi , X) 
\big[
\epsilon_{\delta \eta \mu \nu } F_{\alpha }{}^{\gamma } F^{\delta \eta } R_{\beta }{}^{\mu }{}_{\gamma }{}^{\nu } X 
+ F_{\alpha }{}^{\eta } (\epsilon_{\delta \mu \nu \rho }  F_{\beta }{}^{\mu } R_{\gamma }{}^{\nu }{}_{\eta }{}^{\rho } 
+ R_{\gamma \eta \delta \mu } \tilde{F}_{\beta }{}^{\mu } 
)
\phi {}^{\gamma } \phi {}^{\delta }
\big]
\phi {}^{\alpha } \phi {}^{\beta }\,.
\end{split}
\end{align}
Thus, from the 17 coupling functions in \eqref{eq: po starting action}, we are left with five: $p_3, p_5, p_9, p_{13}, p_{16}$. As we will show below, all $L_{p_i}$, except $L_{p_9}$, vanish identically or reduce to total derivatives.

We start with $p_5$. Due to the diffeomorphism invariance of the action, the scalar field equation of motion is redundant (as long as $\phi_\mu\neq0$) and, therefore, it is sufficient to check only metric and vector field equations. Taking the variation with respect to the vector field yields
\begin{align}
\frac{\epsilon^{\mu }{}_{\nu \eta \zeta }}{\sqrt{-g}}\, \frac{\delta S_{p_5}}{\delta A^{\mu}}= 3!\, p_{5}{}(\phi , X)
g_{\alpha[\eta} \phi {}_{|\beta|}{}_{\nu }{}_{\zeta ]} \phi {}^{\alpha } \phi {}^{\beta }  = 0\,.
\end{align}
Taking variation with respect to the metric, sorting covariant derivatives and using Bianchi identity, it is straightforward to show that 
\begin{align}
\begin{split}
&\frac{\epsilon^{\mu}_{\;\;\xi \eta \zeta }}{\sqrt{-g}}\, \frac{\delta S_{p_5}}{\delta g^{\mu\nu}}
= \frac{4!}{4} p_{5}{}(\phi , X) A^{\mu } \phi_\nu
R_{[\zeta \xi \eta |\mu| } \phi {}_{\alpha ]}
\phi^\alpha = 0\;.
\end{split}
\end{align}
The above analysis proves that $L_{p_5}$ is a total derivative and we can set it to zero
\begin{align}\label{p-zero-5}
\boxed{p_{5}(\phi,X) = 0 \,.}
\end{align}

Similarly, using the dimension-dependent identities, one can show that $L_{p_{13}}$ and $L_{p_{16}}$ are identically zero:
\begin{align*}
    L_{p_{13}} &= \frac{5!}{192} p_{13} F^{\gamma \delta } F^{\rho \sigma } g^{\zeta \eta } g^{\xi \chi } R_{\gamma \delta \xi \zeta } \phi {}^{\alpha } 
    \epsilon_{[\rho \sigma \chi \eta } \phi {}_{\alpha]} = 0\;,
    \\
    L_{p_{16}} &= -\frac{5!}{48} p_{16} F_{\delta }{}^{\omega } F^{\sigma \chi } R^{\gamma }{}_{\omega }{}^{\eta }{}_{\zeta } \phi {}^{\beta } \phi {}^{\delta } \phi {}^{\zeta }
    \epsilon_{[\chi \sigma \gamma \eta } \phi {}_{\beta] } = 0\;.
 \end{align*}
 We thus set
\begin{align}\label{p-zero-13-16}
\boxed{p_{13}(\phi,X) = 0 \,,
	\qquad
	p_{16}(\phi,X) = 0 \,.}
\end{align}

Finally, one can show that $L_{p_3}$ is also identically zero such that we can set
\begin{align}\label{p-zero-3}
\boxed{p_3 (\phi,X) = 0 \,.}
\end{align}
However, this task is much more involved. Indeed, we could only prove it by using the \textit{xAct SymSpin} package~\cite{Aksteiner:2022fmf} for \textit{Mathematica} which implements the spin representation and decomposing the tensors into the irreducible representation.\footnote{\href{https://github.com/Furton/-Decomposing-the-tensors-into-the-irreducible-representation}{https://github.com/Furton/-Decomposing-the-tensors-into-the-irreducible-representation.}}

Thus, for the parity-odd part of the Lagrangian only $L_{p_9}$, which is defined in \eqref{L-p_i-def} left. We have explicitly confirmed that $L_{p_9}$ generates non-trivial dynamics. However, as we will show, it can be completely expressed in terms of the lower-order action and therefore it is redundant. To this end, we obtain the condition under which the whole parity-odd Lagrangian becomes a total derivative. The total parity-odd Lagrangian, after imposing \eqref{p-zero-5}, \eqref{p-zero-13-16}, \eqref{p-zero-3}, and then adding $g_1$, that is defined in \eqref{L-po-EMS}, is given by
\begin{align}\label{L-po-HOMES}
{L}^{{\rm HO}, po}_{\rm MES} = g_1(\phi,X)\, Y + L_{p_{9}} \;,
\end{align}
where the explicit form of $L_{p_{9}}$ is 
shown in \eqref{L-p_i-def}. Choosing the field configuration \eqref{eq: field configurations}, we find that the necessary condition under which the total parity-odd Lagrangian \eqref{L-po-HOMES} becomes a total derivative is
\begin{align}
\label{req: pv Lagrangian is total der}
g_1(\phi, X) =  c_1 + \tfrac{1}{2} X p_{9\phi} \;.
\end{align}
Note that the constant $c_1$ is irrelevant for our purposes, since $c_1Y = c_1 F_{\mu\nu} \tilde{F^{\mu\nu}}$ is a total derivative for any value of $c_1$ and does not contribute to the equations of motion. Eq.~\eqref{req: pv Lagrangian is total der} determines the necessary condition. In order to check whether this is also sufficient condition, we need to substitute \eqref{req: pv Lagrangian is total der} into the covariant equations of motion to see whether they automatically satisfy or not. Substituting \eqref{req: pv Lagrangian is total der} into the vector equations of motion, it is straightforward to show that 
\begin{align}
\frac{\epsilon^{\mu }{}_{\eta \zeta \sigma }}{\sqrt{-g}} \frac{\delta S^{po}}{\delta A^{\mu}} 
&= 3
p_{9 \phi} A^{\mu } \left(
2 R_{[\eta |\mu| \sigma \alpha } \phi_{\zeta]}
+ R_{[\eta \zeta |\mu\alpha|} \phi_{\sigma]}
\right) \phi^{\alpha} = 0\;.
\end{align}
Taking variation with respect to metric, gives
\begin{align}
\label{eq: eomg for Spo}
&\frac{\epsilon^{\mu }{}_{\eta \zeta \sigma }}{\sqrt{-g}} \frac{\delta S^{po}}{\delta g^{\mu\nu}} 
= - \frac{1}{32} \left[ p_{9} \left( I^{(1)}_1 - 2 I^{(1)}_5 \right) 
+  p_{9 \phi}  I^{(1)}_2
+ p_{9 X} \left( I^{(1)}_3 - I^{(1)}_4 \right) \right] = 0
\;,
\end{align}
since each of $I^{(1)}_i$ vanishes separately
\begin{align*}
    I^{(1)}_1 &\equiv  5!\,
    g^{\mu \rho } F_{[\eta \sigma } F_{\nu \mu } \phi {}_{\zeta] }{}_{\rho } = 0 \;,\\
    I^{(1)}_2 &\equiv  5!\, \phi {}^{\mu }
    F_{[\eta \sigma } F_{\nu \mu } \phi {}_{\zeta] } = 0 \;,\\
    I^{(1)}_3 &\equiv  5!\, \phi {}^{\alpha } \phi {}^{\mu }
    F_{[\zeta \sigma } F_{\nu \mu } \phi {}_{\eta] }{}_{\alpha } = 0 \;,\\
    I^{(1)}_4 &\equiv 5!\, \phi^{\alpha} \phi_{\alpha}{}^{\beta} \phi {}^{\mu }
    F_{[\eta \sigma } F_{\mu \beta } g_{\zeta] \nu } = 0 \;,\\
    I^{(1)}_5 &\equiv 5!\, g^{\alpha \beta } \phi^{\mu} F_{[\sigma\mu}\nabla_{|\alpha|}F_{\zeta\beta} g_{\eta]\nu} = 0 \;.
\end{align*}
Here, for clarity we omit indices for $I^{(1)}_{i}$. The expression \eqref{eq: eomg for Spo} is zero, because it can be decomposed as a sum of 5 dimension-dependent identities, each of which is \textit{a priori} zero. Therefore, we have shown that the equations of motion for metric and vector field vanish. As a consequence of diffeomorphism invariance, the scalar field equation should also vanish. Hence, condition \eqref{req: pv Lagrangian is total der} is not only necessary but also sufficient and we can set
\begin{align}\label{p-zero-0}
\boxed{p_{9}(\phi,X) = 0 \,.}
\end{align}

The above results show that there is no parity-violating higher-derivative term in the first subclass of HOMES theory defined by conditions $\textbf{I-1}$ to $\textbf{I-4}$ in Sec.~\ref{sec: Introduction}. Thus, in the odd-sector, the most general Lagrangian is given by \eqref{L-po-EMS}. This completes the analysis leading to the result quoted in Sec.~\ref{sec: The HOMES Lagrangian}.

\section{Lagrangian construction for the second subclass}
\label{sec: Lagrangian construction for the second case}

In this section, we study the second subclass that is defined by conditions $\textbf{II-1}$ to $\textbf{II-3}$ in Sec.~\ref{sec: Introduction} and we show that the most general form for the Lagrangian is given by Eq.~\eqref{L-pe-EMS} with $w_0=0$.

For the second subclass of HOMES, the Lagrangian should have the general form 
\begin{align}\label{L-2nd-subclass-def}
{L}^{{\rm HO}}_{\rm MES} &=\sum_{i} P_i(\phi,X,F,Z,Y)  T_i^{\mu\nu} \phi_{\mu\nu} \,,
\end{align}
where $P_i$ are unspecified functions of scalar field and variables that are defined in \eqref{eq: variables FXYZ}. 

We first need to find the explicit form of the symmetric rank-2 tensor $T_i^{\mu\nu}$. Without scalar field derivative $\phi^\mu$, they are given by $g^{\mu\nu}$ and $\tilde{F}^{\mu \rho } \tilde{F}_{\rho }{}^{\nu }$. Note that we do not need to include terms $F^{\mu \rho } F_{\rho }{}^{\nu }$ and $F^{\mu \rho } \tilde{F}_{\rho }{}^{\nu }$ since they are redundant due to the identities \eqref{eq: identities}. In order to find the terms which include $\phi^\mu$, it is more convenient to define the symmetric tensors $B_{IJ}^{\mu\nu}$ as follows
\begin{align}
B_{IJ}^{\mu\nu} \equiv \frac{1}{2} \left(B^\mu_I B^\nu_J + B^\nu_I B^\mu_J \right) \,,
\end{align}
where 
\begin{align}
B_I^\mu = \{ 
B_0^\mu, B_1^{\mu}, \tilde{B}_1^\mu, B_2^{\mu}
\}\,;
\qquad
I = 0,1,\tilde{1},2\,,
\end{align}
are defined in Eq.~\eqref{B-def}. For example $B_{00}^{\mu\nu} = B_0^{\mu } B_0^{\nu }$, $B_{01}^{\mu\nu} = \tfrac{1}{2} (B_1^{\nu } B_0^{\mu } + B_1^{\mu } B_0^{\nu })$ and so on. The power of vector field is determined by $I+J$ such that $B_{00}^{\mu\nu}$ does not include any vector field while $B_{\tilde{1}2}^{\mu\nu}$ includes third power of vector field. Note that not all $B_{IJ}^{\mu\nu}$ are independent building blocks. For example, we do not need to include the term $B_{11}^{\mu\nu}$ since it is redundant
$B_{11}^{\mu\nu} 
= Z g^{\mu \nu } 
+2 X \tilde{F}^{\mu \rho } \tilde{F}_{\rho }{}^{\nu }
- 2 F B_{00}^{\mu\nu}
- B_{\tilde{1}\tilde{1}}^{\mu\nu}
+ 2 B_{02}^{\mu\nu}$. Taking into account all independent $B_{IJ}^{\mu\nu}$ together with the two terms without scalar field derivative $g^{\mu\nu}$ and $\tilde{F}^{\mu \rho } \tilde{F}_{\rho }{}^{\nu }$, the independent $T_i^{\mu\nu}$ turns out to be
\begin{align}\label{T-i-def}
T_i^{\mu\nu} = 
\{
g^{\mu\nu}, 
\tilde{F}^{\mu \rho } \tilde{F}_{\rho }{}^{\nu },
B_{00}^{\mu\nu}, B_{01}^{\mu\nu}, B_{0\tilde{1}}^{\mu\nu}, B_{02}^{\mu\nu}, B_{1\tilde{1}}^{\mu\nu}, B_{12}^{\mu\nu}, B_{\tilde{1}\tilde{1}}^{\mu\nu}, B_{\tilde{1}2}^{\mu\nu}, B_{22}^{\mu\nu}
\}\,; 
\qquad
i = 1,\cdot\cdot,11 \,.
\end{align}

Substituting \eqref{T-i-def} in \eqref{L-2nd-subclass-def}, one obtains the explicit form of the Lagrangian. Following the same reasoning as in the previous sections, we will begin by obtaining the necessary conditions for the absence of higher derivatives in the equations of motion generated by the  Lagrangian. Therefore, we take the variation with respect to the vector field, then substituting the following field configurations
\begin{align}
\begin{split}
g_{\mu\nu} &=\text{diag} (-1,a_1^2(t,x),a_2^2(t,x),a_3^2(t,x)),\;\\
A_{\mu} &= (0,A_1, A_2, A_3),\; \qquad  \phi \,,
\end{split}
\end{align}
with the following six choices
\begin{align}\label{configs}
\begin{split}
(1):\, A_1 &= A_1(t)\;, \qquad A_2 = A_2(x),\; \qquad A_3 = A_3(t),\; \qquad \phi = \phi(t)\,;\;\\
(2):\, A_1 &= A_1(t)\;, \qquad A_2 = A_2(t),\; \qquad A_3  = A_3 (y),\;\qquad \phi = \phi(x)\,;\;\\
(3):\, A_1 &= A_1(t)\;, \qquad A_2 = A_2(x),\; \qquad A_3  = A_3 (y),\; \qquad \phi = \phi(x)\,;\;\\
(4):\, A_1 &= A_1(t)\;,\qquad A_2 = A_2(x),\; \qquad A_3  = A_3 (t,y),\; \quad \phi = \phi(x)\,;\;\\
(5):\, A_1 &= A_1(t,x)\;, \quad A_2 = A_2(t,x),\; \quad A_3  = A_3 (t),\;\qquad \phi = \phi(x)\,;\\
(6):\, A_1 &= A_1(t)\;, \qquad A_2 = A_2(x),\;\qquad A_3  = A_3 (t,y),\;\quad \phi = \phi(t,x)\,,
\end{split}
\end{align}
one after another. More precisely, we start with configuration 1 above. We substitute it in the covariant equation for the vector field and by requiring the absence of higher-order derivatives, we find some conditions. Substituting back these conditions into the covariant equation for the vector field, we try the next configuration and we continue this procedure until the last configuration in \eqref{configs}. We find that the covariant equation of motion becomes second-order at this stage and we terminate the procedure. This process leads to the set of necessary and sufficient conditions which after substituting back into the Lagrangian yields
\begin{align}\label{L-c}
\begin{split}
{L}^{{\rm HO}}\big|_{\text{n.c.}} 
&= s_{9}{}(\phi , X) \square\phi 
+ s_{16}{}(\phi , X) \phi {}^{\alpha } \phi {}^{\beta } \phi {}_{\alpha }{}_{\beta } 
- 2 s_{10}{}(\phi , X) \tilde{F}_{\alpha }{}^{\gamma } \phi {}^{\alpha } \phi {}^{\beta } \phi {}_{\beta }{}_{\gamma } 
\\
&+ s_{12}{}(\phi , X) \tilde{F}_{\alpha }{}^{\gamma } \tilde{F}_{\beta }{}^{\delta } \phi {}^{\alpha } \phi {}^{\beta } \phi {}_{\gamma }{}_{\delta } 
+ s_{15}{}(\phi , X) \tilde{F}_{\alpha }{}^{\gamma } \tilde{F}^{\alpha \beta } \phi {}_{\beta }{}_{\gamma } 
+ P_{6}{}(\phi , X, F, Z, Y) C_6 \;,
\end{split}
\end{align}
where
\begin{align*}
C_6 &\equiv - \tfrac{1}{2} X Y^2 \square\phi 
- \left(16 F^2 + \tfrac{1}{4} Y^2 \right) \phi {}^{\alpha } \phi {}_{\alpha }{}_{\beta } \phi {}^{\beta } 
+ 4 (Z-4XF) \tilde{F}_{\alpha }{}^{\gamma } \tilde{F}^{\alpha \beta } \phi {}_{\beta}{}_{\gamma }
\\
&+ 2 \big( 
8 F F_{\alpha }{}^{\gamma } F_{\gamma }{}^{\delta } \phi {}_{\beta }{}_{\delta } 
- 4 F \tilde{F}_{\alpha }{}^{\gamma } \tilde{F}_{\beta }{}^{\delta } \phi {}_{\gamma}{}_{\delta } 
- Y F_{\alpha }{}^{\gamma } \tilde{F}_{\beta }{}^{\delta } \phi{}_{\gamma }{}_{\delta } 
- 2 F_{\alpha }{}^{\gamma } F_{\beta }{}^{\delta } F_{\gamma }{}^{\zeta } F_{\delta }{}^{\eta } \phi {}_{\zeta }{}_{\eta }
\big) 
\phi {}^{\alpha } \phi {}^{\beta } \,.
\end{align*}
While the configuration \eqref{configs} is not unique and there are many other possibilities, the Lagrangian \eqref{L-c} is unique. This is enough for our purpose. Indeed, a careful choice and ordering of background configurations is key to be able to quickly derive the conditions that ensure the absence of higher-derivative terms in the covariant equations of motion. 

Imposing second-order equations of motion, reduces 11 terms in Lagrangian \eqref{L-2nd-subclass-def} to 6 term in \eqref{L-c}. The $s_9$ and $s_{16}$ terms are nothing but the scalar coupling $s_1$ and $s_3$ in \eqref{scalar-couplings} which are not independent and give the KGB term. The $s_{15}$ and $s_{12}$ terms are $w_1$ and $w_2$ that were already found in the first subclass. Variation of $s_{10}$ term with respect to scalar field, vector field and metric vanish and, therefore, we can set it to zero
\begin{align}\label{s10-zero-0}
\boxed{s_{10}(\phi,X) = 0 \,.}
\end{align}
We thus only need to focus on $P_6$ term. 

Indeed, the operator $C_6$ can be written as a sum of four dimension-dependent identities
\begin{align*}
C_6 = -\frac{45}{4} \left( 2 I^{(2)}_{1} + I^{(2)}_{2} + I^{(2)}_{3} \right) - 60 I^{(2)}_{4} \;,
\end{align*}
where
\begin{align*}
I^{(2)}_{1} &\equiv  F_{[\gamma \zeta } F_{\delta \chi }  \phi {}_{\beta]  } \phi {}_{\alpha }{}^{\beta } F^\zeta{}_{\eta } F^{\gamma \delta } g^{\eta\chi } \phi {}^{\alpha }  = 0 \;,
\\
I^{(2)}_{2} &\equiv F_{[\beta \gamma } F_{\delta \zeta } \phi {}_{\alpha] } \phi {}_{\eta }{}_{\chi } F^{\beta \gamma } F^{\delta \zeta } g^{\eta \chi } \phi {}^{\alpha } = 0 \;,
\\
I^{(2)}_{3} &\equiv F_{[\beta \gamma } F_{\delta \zeta }  \phi {}_{\eta ]}{}_{\chi } \phi {}_{\alpha } F^{\beta \gamma } F^{\delta \zeta } g^{\eta \chi } \phi {}^{\alpha } = 0 \;,
\\
I^{(2)}_{4} &\equiv F_{\gamma [\beta } F_{\delta \zeta } \phi {}_{\alpha } \phi {}_{\eta] }{}_{\chi } F^{\beta \gamma } F^{\delta \zeta } g^{\eta \chi } \phi {}^{\alpha } = 0 \;,
\end{align*}
which shows that $C_6$ identically vanishes and we can set
\begin{align}\label{P6-zero-0}
\boxed{P_{6}(\phi,X,F,Z,Y) = 0 \,.}
\end{align}

Therefore, we conclude that the Lagrangian for the second subclass does not lead to any new higher-derivative terms. In summary, the second subclass leads to
\begin{align*}
L_{\rm H} &= f(X,F,Y,X) - G_3(\phi,X)\,\square\phi 
+ \big[
 w_1(\phi,X) g_{\beta\delta} 
+ w_2(\phi,X) \phi_{\beta} \phi_{\delta}  
\big] \phi_{\gamma\alpha} \tilde{F}^{\alpha \beta } \tilde{F}^{\gamma \delta } \;, 
\end{align*} 
where $f$ is an arbitrary function. Note that the higher-order part of the second subclass, defined by conditions $\textbf{II-1}$ to $\textbf{II-3}$ in Sec.~\ref{sec: Introduction}, coincides with that of the first subclass, defined by conditions $\textbf{I-1}$ to $\textbf{I-4}$, when the Horndeski non-minimal term is set to zero, $w_0=0$. This completes our proof.

\section{Summary and discussions}
\label{sec: Conclusion}

We have studied the Higher-Order Maxwell–Einstein–Scalar (HOMES) theories, which involves a vector field and a scalar field, and respects spacetime diffeomorphism and $U(1)$ gauge symmetries. We have focused on the linear HOMES theories, which includes terms up to the first order in second derivative of all fields, namely $\{\phi_{\mu\nu},\nabla_\rho{F}_{\mu\nu}, R_{\mu\nu\alpha\beta}\}$. In the absence of the vector field, the linear HOMES theories reduce to the ${\rm KGB}$ scalar-tensor theory \cite{Deffayet:2010qz} which includes terms up to the linear order in $\{\phi_{\mu\nu}, R_{\mu\nu\alpha\beta}\}$. In this sense, HOMES can be regarded as a natural generalization of KGB which includes a $U(1)$ vector field. Furthermore, although HOMES theories provide the same number of degrees of freedom as generalized Proca theories \cite{Tasinato:2014eka,Heisenberg:2014rta}, they are fundamentally distinct due to the presence of $U(1)$ gauge symmetry in the vector sector.

We have considered two broad subclasses of linear HOMES theories. The first subclass, that is defined by conditions $\textbf{I-1}$ to $\textbf{I-4}$ in Sec.~\ref{sec: Introduction}, includes terms in the action that are up to quadratic in $F_{\mu\nu}$ and contains terms up to linear order in the second derivatives of all fields, namely $\{\phi_{\mu\nu},\nabla_\rho{F}_{\mu\nu}, R_{\mu\nu\alpha\beta}\}$. In general, such an action leads to higher-order (beyond second-order) equations of motion typically resulting in the propagation of extra degrees of freedom, associated with the so-called Ostrogradsky ghost. By taking into account all symmetries, we have found $24$ and $17$ independent terms for the Lagrangians in the parity-even and parity-odd sectors respectively. Then, imposing the necessary and sufficient conditions that ensure second-order equations of motion for all fields, we have found the most general Lagrangians, given by Eqs.~\eqref{L-pe-EMS} and \eqref{L-po-EMS}. There are four independent higher-derivative terms in the parity-even sector Eq.~\eqref{L-pe-EMS}: KGB term $G_3(\phi,X)\Box\phi$ in the scalar sector, the Horndeski non-minimal term $w_0(\phi)R_{ \beta \delta \alpha \gamma}\tilde{F}^{\alpha \beta } \tilde{F}^{\gamma \delta }$ in the vector field sector, and two interaction terms between the scalar and vector field sectors 
$[w_1(\phi,X) g_{\rho\sigma} + w_2(\phi,X) \phi_{\rho} \phi_{\sigma}] \phi_{\beta\alpha}  \tilde{F}^{\alpha \rho } \tilde{F}^{\beta\sigma}$. We have found that there is no higher-derivative term in the parity-odd sector as shown in Eq.\eqref{L-po-EMS}.

We have then considered the second subclass of linear HOMES theories that is defined by conditions $\textbf{II-1}$ to $\textbf{II-3}$ in Sec.~\ref{sec: Introduction}. This subclass includes terms up to linear in $\phi_{\mu\nu}$ but does not contain $\{\nabla_\rho{F}_{\mu\nu}, R_{\mu\nu\alpha\beta}\}$. Unlike the first subclass, however, we allow for arbitrary powers/functions of $F_{\mu\nu}$. In other words, the Lagrangian now contains arbitrary functions of scalar field, kinetic term of scalar field, and all possible non-redundant contractions of the field strength tensor and/or first derivative of the scalar field. We have found that there are $11$ independent terms in this case. Imposing the second-order equations of motion, we have found that the resulting theory admits only three of the four higher-derivative terms that are obtained for the first subclass. Thus, for both subclasses, we explicitly demonstrate that \textit{no} further generalization is possible under the stated conditions: one is inevitably left with exactly \textit{four} higher-derivative terms for the first subclass and \textit{three} higher-derivative terms for the second subclass.

In order to go beyond these higher-order terms in the context of linear HOMES theories, we have considered the case when the scalar field is complex and it is charged under the $U(1)$ gauge symmetry. We have presented explicit examples of this scenario in Eq. \eqref{L-complex} with new higher-derivative interaction terms which generate second-order equations of motion for all fields. Interestingly, while only two higher-order mixing terms are allowed in the case of a real scalar field, there are three interactions terms between the scalar and vector field sectors for a complex scalar field.

We emphasize that HOMES theories are different from the generalized Proca theories as the first includes second (or higher) derivatives of the vector field while latter does not. It would be interesting to extend the present analysis to the case of quadratic or higher-order HOMES theories and also perform a systematic study of the complex scalar case. One could also consider the Degenerate Higher-Order Maxwell–Einstein–Scalar (DHOMES) theories or/and the Unitary-gauge Higher-Order Maxwell–Einstein–Scalar (U-DHOMES) theories. We leave these rich and promising directions for future works.

\vspace{0.5cm}

{\bf Acknowledgments:} 
We acknowledge the use of the xTras package \cite{Nutma:2013zea} for tensorial calculations. We are grateful to Thomas B\"{a}ckdahl  for his help with the \textit{xAct SymSpin} package, and to Sebastian Bahamonde, Lavinia Heisenberg and Sergey Mironov for useful discussions. The work of M.A.G., P.P., and M.Y. is supported by IBS under the project code, IBS-R018-D3. The work of S.M. is supported in part by JSPS Grant-in-Aid for Scientific Research Number~JP24K07017 and the World Premier International Research Center Initiative (WPI), MEXT, Japan. M.Y. is supported in part by JSPS Grant-in-Aid for Scientific Research Number JP23K20843. 

\vspace{0.0cm}

\appendix

\numberwithin{equation}{section}

\bibliographystyle{JHEPmod}
\bibliography{ref}

\end{document}